\documentclass[nojss]{jss}

\usepackage[utf8]{inputenc}
\usepackage[english]{babel}

\usepackage{amsmath}
\usepackage{amsfonts}

\newcommand{\class}[1]{\mbox{\textsf{#1}}}
\newcommand{\func}[1]{\mbox{\texttt{#1()}}}
\newcommand{\set}[1]{\mathcal{#1}}
\newcommand{\mat}[1]{{\mathbf{#1}}}
\author{Michael Hahsler}

\title{\pkg{recommenderlab}: An R Framework for Developing and Testing Recommendation Algorithms}

\Plainauthor{Michael Hahsler} 
\Plaintitle{recommenderlab: An R Framework for Developing and Testing Recommendation Algorithms} 
\Shorttitle{Developing and Testing Recommendation Algorithms with R} 

\Abstract{Algorithms that create recommendations based on observed data have
significant commercial value for online retailers and many other industries.
Recommender systems has a significant research
community and studying such systems is part of most modern data science
curricula. While there is an abundance of software that implements recommendation algorithms, there is little in terms of supporting recommender system research and education. This paper describes the open-source software~\pkg{recommenderlab} which was created with supporting research and education in mind.
The package can be directly installed in R or downloaded from
\url{https://github.com/mhahsler/recommenderlab}.}

\Keywords{recommender systems, collaborative filtering, evaluation}
\Plainkeywords{recommender systems, collaborative filtering, evaluation} 

\Address{
  Michael Hahsler\\
	Computer Science\\
	Lyle School of Engineering\\
	Southern Methodist University\\
	P.O. Box 750122 \\
	Dallas, TX 75275-0122\\
	E-mail: \email{mhahsler@lyle.smu.edu}\\
	URL: \url{michael@hahsler.net}
}

\begin{document}



\section{Introduction}
Recommender
systems apply statistical and knowledge discovery techniques to the problem of
making product recommendations based on previously recorded
usage data~\citep{recommender:Sarwar:2000}.
Creating such automatically generated personalized recommendations for
products including
books, songs, TV shows and movies using collaborative filtering
have come a long way since
Information Lense,
the first system using
\emph{social filtering} was created
more than 30 years ago~\citep{recommender:Malone:1987}.
Today, recommender systems
are a successful technology used by market
leaders in several industries (e.g., by
Amazon, 
Netflix, and 
Pandora). 
In retail, such
recommendations can improve conversion rates by helping
the customer to find
products she/he wants to buy faster, promote cross-selling by suggesting
additional products and can improve customer loyalty through creating a
value-added relationship~\citep{recommender:Schafer:2001}.

Even after 30 years, recommender systems still have a very active research
community.
It is often not clear which of the many available algorithms is appropriate for a particular application and new approaches are constantly proposed. Many commercially available software applications implement recommender
algorithms, however,
this paper focuses on software support for recommender systems
research which includes rapid prototyping algorithms and thorough
evaluation and comparison of algorithms.
For this purpose, access to the source code is paramount.
Many open-source projects implementing recommender algorithms have been
initiated over the years. Table~\ref{tab:software} provides links to several popular
open source implementations
which provide code which can be used by researchers.
The extent of (currently available) functionality as well as the target
usage of the available software packages vary greatly and many projects have been abandoned over the years. A comprehensive list of recommender systems software is maintained by~\cite{Jenson:2019}.


\begin{table}
\begin{tabular}{|p{1.9cm}|p{3cm}|l|l|}
\hline
{\bf Software} & {\bf Description} & {\bf Language} & {\bf URL} \\
\hline
Apache Mahout & Machine learning library includes collaborative filtering &Java& \small\url{http://mahout.apache.org/}\\
\hline
Crab & Components to create recommender systems & Python & \small\url{https://github.com/muricoca/crab}\\
\hline
LensKit & Collaborative filtering algorithms from GroupLens Research & Python & \small\url{http://lenskit.grouplens.org/}\\
\hline
MyMediaLite & Recommender system algorithms. & C\#/Mono & \small\url{http://www.mymedialite.net/}\\
\hline
SVDFeature & Toolkit for feature-based matrix factorization& C++ & \small\url{https://www.jmlr.org/papers/v13/chen12a.html}\\
\hline
Vogoo PHP LIB & Collaborative filtering engine for personalizing web sites & PHP & \small\url{http://sourceforge.net/projects/vogoo/} \\

\hline
\end{tabular}
\caption{Recommender System Software freely available for research.}
\label{tab:software}
\end{table}

Most available software focuses on creating
recommender applications for deployment
as a production system or they implement a single method as part of a research
project.
The R extension package~\pkg{recommenderlab}
described in this paper
was designed for a completely different purpose.
It aims at providing a comprehensive research infrastructure for
recommender systems.
The focus is on consistent and efficient data handling,
easy incorporation of existing and new algorithms,
experiment set up and evaluation of the results.
The open-source programming language R, a popular software
environment for statistical computing and data scientists~\citep{R:2018},
is used as the platform since it easily allows the researcher to
either implement and integrate algorithms written in a wide range of programming languages including R, Python, Java, C/C++
and already provides all needed statistical tools
which is important to provide a useful research environment.

Although
developed to support the author's own research and teaching needs
in 2010,
there proved to be a need for research-focused software and the
package \pkg{recommenderlab} turned out to be quite popular.
A healthy community of users who file bug reports, suggest improvements and contribute their own algorithms has grown around the package and is coordinated using the software's GitHub page\footnote{\url{https://github.com/mhahsler/recommenderlab}}.
Authors not related to the developer have written a textbook about how to use the package~\citep{Suresh:2015}. The package is used in several university courses to demonstrate the basics of recommender system development.
Finally, the package was employed by several researchers to develop and test their own algorithms~\citep[e.g.,][]{Chen:2013, Buhl:2016, Beel:2016, Lombardi:2017}.

Package~\pkg{recommenderlab} focuses on collaborative filtering
which is based on the idea that given rating data by many users for many items
(e.g., 1 to 5 stars),
one can predict a user's rating for an item
not known to her or him~\citep[see, e.g.,][]{recommender:Goldberg:1992}
or create
for each user a so called top-$N$ lists of recommended
items~\citep[see, e.g.,][]{recommender:Sarwar:2001,recommender:Deshpande:2004}.
The premise is that users who agreed on the rating for some items
typically also tend to agree on the rating for other items.

\pkg{recommenderlab} provides implementations of many popular algorithms, including the following.

\begin{itemize}
    \item {\bf User-based collaborative filtering (UBCF)} predicts ratings by aggregating the ratings of users who have a similar rating history to the active user~\citep{recommender:Goldberg:1992,Resnick:1994,recommender:Shardanand:1995}.
    \item {\bf Item-based collaborative filtering (IBCF)} uses item-to-item similarity based on user ratings to find items that are similar to the items the active user likes~\citep{Kitts:2000,recommender:Sarwar:2001,
recommender:Linden:2003,recommender:Deshpande:2004}.
    \item {\bf Latent factor models} use singular value decomposition (SVD) to
        estimate missing ratings using methods like SVD with column-mean imputation, Funk SVD or alternating least squares~\citep{Hu:2008,recommender:Koren:2009}.
    \item {\bf Association rule-based recommender (AR)} uses association rules to find recommended items~\citep{Fu:2000,Mobasher:2001,Geyer-Schulz:2002,Lin:2002,Demiriz:2004}.
    \item {\bf Popular items (POPULAR)} is a non-personalized algorithm which recommends to all users the most popular items they have not rated yet.
    \item {\bf Randomly chosen items (RANDOM)} creates random recommendations which can be used as a baseline for recommender algorithm evaluation.
    \item {\bf Re-recommend liked items (RERECOMMEND)} recommends items which the user has rated highly in the past. These recommendations can be useful for items that are typically consumed more than once (e.g., listening to songs or buying groceries).
    \item {\bf Hybrid recommendations (HybridRecommender)} aggregates the recommendations of several algorithms~\citep{Cano:2017}.
\end{itemize}

We will discuss some of these algorithms in the rest of the paper. Detailed information
can be found in the
survey book by~\cite{recommender:Desrosiers:2011}.

This rest of this paper is structured as follows. Section~\ref{sec:CF} introduces
collaborative filtering and
some of its popular algorithms.
In section~\ref{sec:evaluation} we discuss the evaluation of recommender
algorithms.
We introduce the infrastructure provided by \pkg{recommenderlab}
in section~\ref{sec:infrastructure}. In section~\ref{sec:examples} we
illustrate the capabilities on the package to create and evaluate
recommender algorithms. We conclude with section~\ref{sec:conclusion}.

\section{Collaborative Filtering}
\label{sec:CF}

To understand the use of the software, a few formal definitions are
necessary.
We will often give examples for a movie recommender, but the examples
generalize to other types of items as well.
Let $\set{U} = \{u_1, u_2, \ldots, u_m\}$ be the set of users
and $\set{I} = \{i_1, i_2, \ldots, i_n\}$ the set of items.
Ratings are stored in a $m \times n$ user-item rating matrix $\mat{R} = (r_{jk})$, where
$r_{jk}$ represents the rating of user
$u_j$ for item $i_k$.
Typically, only a small fraction of ratings are
known and for many cells in $\mat{R}$, the values are missing.
Missing values represent movies that the user has not rated and potentially also not seen yet.

Collaborative filtering aims to create recommendations for a user
called the active user $u_a \in \set{U}$.
We define the set of items unknown to user $u_a$ as
$\set{I}_a = \set{I} \setminus \{i_l \in \set{I}| r_{al}\ \textrm{is not missing}\}$.
The two typical tasks are to predict
ratings for all items in $\set{I}_a$ or to create a list containing the best $N$ recommended items from $\set{I}_a$
(i.e., a top-$N$ recommendation list) for
$u_a$.
Predicting all missing ratings means completing the row of the
rating matrix
where the missing values
for items in $\set{I}_a$
are replaced by ratings estimated from other data in $\mat{R}$.
From this point of view, recommender systems are related to matrix completion
problem.
Creating a top-$N$ list can
be seen as a second step after predicting ratings for all
unknown items in $\set{I}_a$
and then taking the $N$ items with the highest predicted ratings.
Some algorithms skip predicting ratings first and are able to
find the top $N$ items directly.
A list
of top-$N$ recommendations for a user $u_a$ is an partially ordered set
$T_N = (\set{X}, \ge)$, where
$\set{X} \subset \set{I}_a$ and
$|\set{X}| \le N$ ($|\cdot|$ denotes the cardinality of the set).
Note that there may exist cases
where top-$N$ lists contain less than $N$ items. This can happen if
$|\set{I}_a| < N$ or if the CF algorithm is unable to
identify $N$ items to recommend.
The binary relation $\ge$ is defined as
$x\ge y$ if and only if
$\hat{r}_{ax} \ge
\hat{r}_{ay}$ for all $x,y \in \set{X}$. Furthermore we
require that $\forall_{x\in \set{X}} \forall_{y\in \set{I}_a} \quad \hat{r}_{ax} \ge \hat{r}_{ay}$ to ensure that the top-$N$ list contains
only the items with the highest estimated rating.

Typically we deal with a very large number of items
with unknown ratings
which makes first predicting rating values for all of them computationally
expensive.
Some approaches (e.g., rule based approaches) can
predict the top-$N$ list directly without considering all unknown items first.

Collaborative filtering algorithms are typically divided into two groups,
\emph{memory-based CF} and \emph{model-based CF}
algorithms~\citep{recommender:Breese:1998}. Memory-based CF
use the whole (or at least a large sample of the) user database to create
recommendations. The most prominent algorithm is
user-based collaborative filtering.
The disadvantages of this approach is scalability since the whole
user database has to be processed online for creating recommendations.
Model-based algorithms
use the user database to learn a more compact model (e.g, clusters
with users of similar preferences) that is later used to create
recommendations.

In the following we will present the basics of well known memory and model-based
collaborative filtering algorithms. Further information about
these algorithms can be found in the recent
survey book chapter by~\cite{recommender:Desrosiers:2011}.

\subsection{User-based Collaborative Filtering}

User-based CF~\citep{recommender:Goldberg:1992,Resnick:1994,recommender:Shardanand:1995} is
a memory-based algorithm which tries to mimics word-of-mouth by analyzing
rating data from many individuals. The assumption is that
users with similar preferences will rate items similarly. Thus
missing ratings for a user can be predicted by first
finding a \emph{neighborhood} of similar users and then aggregate the
ratings of these users to form a prediction.

The neighborhood is defined in terms of similarity between users,
either by taking a given number of most similar users ($k$ nearest neighbors)
or all users within
a given similarity threshold.
Popular similarity measures for CF are
the \emph{Pearson correlation coefficient} and
the \emph{Cosine similarity}. These similarity measures are defined
between two users $u_x$ and $u_y$ as
\begin{equation}
\mathrm{sim_{Pearson}}(\vec{x},\vec{y}) =
	    \frac{\sum_{i \in I} (\vec{x}_i \, \bar{\vec{x}})(\vec{y}_i \, \bar{\vec{y}})}
		{(|I| -1) \, \mathrm{sd}(\vec{x}) \, \mathrm{sd}(\vec{y})}
\end{equation}
and
\begin{equation}
\mathrm{sim_{Cosine}}(\vec{x},\vec{y}) =
	    \frac{\vec{x}\cdot\vec{y}}
		    {\|\vec{x}\| \|\vec{y}\|},
\end{equation}
where
$\vec{x} = r_{x}$ and
$\vec{y} = r_{y}$ represent the
row vectors in $\mat{R}$ with the
two users' profile vectors.
$\mathrm{sd}(\cdot)$ is the standard deviation and
$\|\cdot\|$ is the $l^2$-norm of a vector.
For calculating similarity using rating data only the dimensions (items)
are used which were rated by both users.

Now the neighborhood for the active user $\set{N}(a) \subset \set{U}$
can be selected
by either a threshold on the similarity or by taking the
$k$ nearest neighbors.
Once the users in the neighborhood are found, their ratings are
aggregated to form the predicted rating for the active user.
The easiest form is to just average the ratings in the neighborhood.

\begin{equation}
\hat{r}_{aj}=\frac{1}{|\set{N}(a)|} \sum_{i\in\set{N}(a)} r_{ij}
\label{equ:aggregation1}
\end{equation}

\begin{figure}
\centerline{\includegraphics[width=13cm]{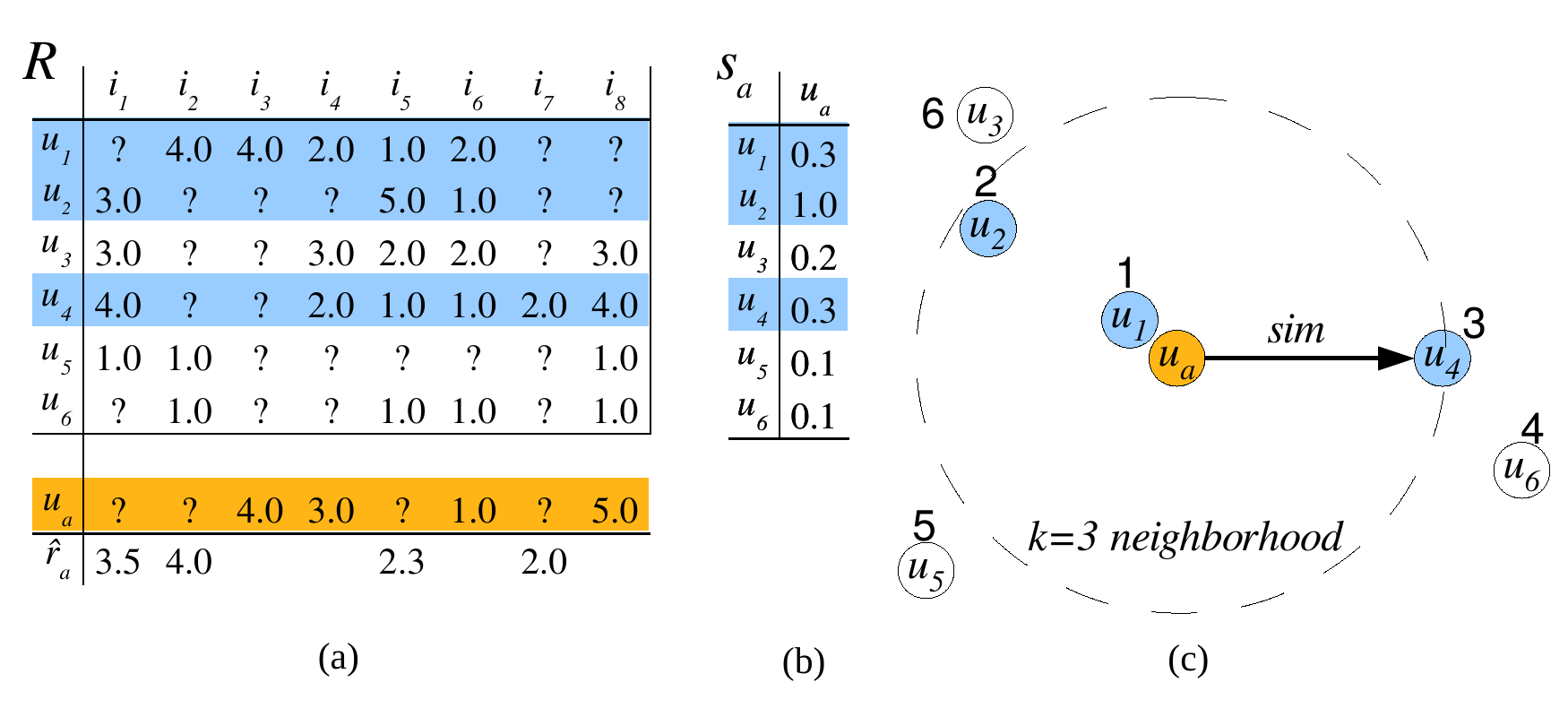}}
\caption{User-based collaborative filtering example with (a) rating matrix $R$
and estimated ratings for the active user, (b), similarites between the active user and the other users $s_a$ (Euclidean distance converted to similarities),
    and
(b) the user neighborhood formation.
}
\label{fig:UBCF}
\end{figure}

An example of the process of creating recommendations by
user-based CF is shown in Figure~\ref{fig:UBCF}. To the left is the rating
matrix $\mat{R}$ with 6 users and 8 items and ratings in the
range 1 to 5 (stars).
We want to create recommendations for the active user $u_a$
shown at the bottom of the matrix.  To find the
$k$-neighborhood (i.e., the $k$ nearest neighbors) we calculate the similarity
between the active user and all other users
based on their ratings
in the database and then select the
$k$ users with the highest similarity. To the right in Figure~\ref{fig:UBCF}
we see a $2$-dimensional representation of the similarities (users with higher
similarity are displayed closer) with the active user in the center.
The $k=3$ nearest
neighbors ($u_1, u_2$ and $u_3$) are selected and marked in the
database to the left. To generate an
aggregated estimated rating, we
compute the average ratings in the neighborhood for each item not
rated by the active user.
To create a top-$N$ recommendation list, the items are ordered by
predicted rating.
In the small example in
Figure~\ref{fig:UBCF} the order in the top-$N$ list (with $N \ge 4$)
is $i_2, i_1, i_7$ and $i_5$. However, for a real application we
probably would
not recommend items $i_7$ and $i_5$ because of their low ratings.

The fact that some users in the neighborhood are more similar to
the active user than others (see Figure~\ref{fig:UBCF} (b)) can be
incorporated as weights into
Equation~(\ref{equ:aggregation1}).

\begin{equation}
\hat{r}_{aj}=\frac{1}{\sum_{i\in\set{N}(a)} s_{ai}} \sum_{i\in\set{N}(a)} s_{ai} r_{ij}
\label{equ:aggregation2}
\end{equation}
$s_{ai}$ is the similarity between the active user~$u_a$ and user
$u_i$ in the neighborhood.

For some types of data the performance of the recommender algorithm can
be improved by removing user rating bias.
This can be done by normalizing the rating data before applying the
recommender algorithm.
Any normalization
function $h: \mathbb{R}^{n \times m} \mapsto \mathbb{R}^{n \times m}$
can be used for preprocessing. Ideally, this function is reversible
to map the predicted rating on the normalized scale back
to the original rating scale.
Normalization is used to remove individual rating bias by
users who consistently always use lower or higher
ratings than other users.
A popular method is to center the rows of the user-item rating matrix
by
$$h(r_{ui}) = r_{ui} - \bar{r}_u,$$
where $\bar{r}_u$ is the mean of all available ratings in row~$u$ of
the user-item rating matrix $\mat{R}$.

Other methods like Z-score normalization which also takes rating variance
into account can be found in the
literature \citep[see, e.g.,][]{recommender:Desrosiers:2011}.

The two main problems of user-based CF are that the whole
user database has to be kept in memory and that
expensive similarity computation between the active user and
all other users in the database has to be performed.

\subsection{Item-based Collaborative Filtering}
Item-based CF~\citep{Kitts:2000,recommender:Sarwar:2001,
recommender:Linden:2003,recommender:Deshpande:2004}
is a model-based approach which produces recommendations
based on the relationship between items inferred from the rating
matrix. The assumption behind this approach is that users will prefer items
that are similar to other items they like.

The model-building step consists of calculating a similarity matrix
containing all item-to-item similarities using a given similarity measure.
Popular are again Pearson correlation and Cosine similarity.
%
All pairwise similarities are stored in a $n \times n$
similarity matrix $\mat{S}$.
To reduce the model size
to $n \times k$ with $k \ll n$,
for each item only a list of the $k$ most similar
items and their similarity values are stored.
The $k$ items which are most similar to item $i$ is
denoted by the set $\set{S}(i)$ which can be seen as the
neighborhood of size $k$ of the item.
Retaining only $k$ similarities per item improves the
space and time complexity significantly but potentially sacrifices
some recommendation quality~\citep{recommender:Sarwar:2001}.

To make a recommendation based on the model
we use the similarities to calculate a weighted sum of the user's ratings
for related items.

\begin{figure}
\centerline{\includegraphics[width=9cm]{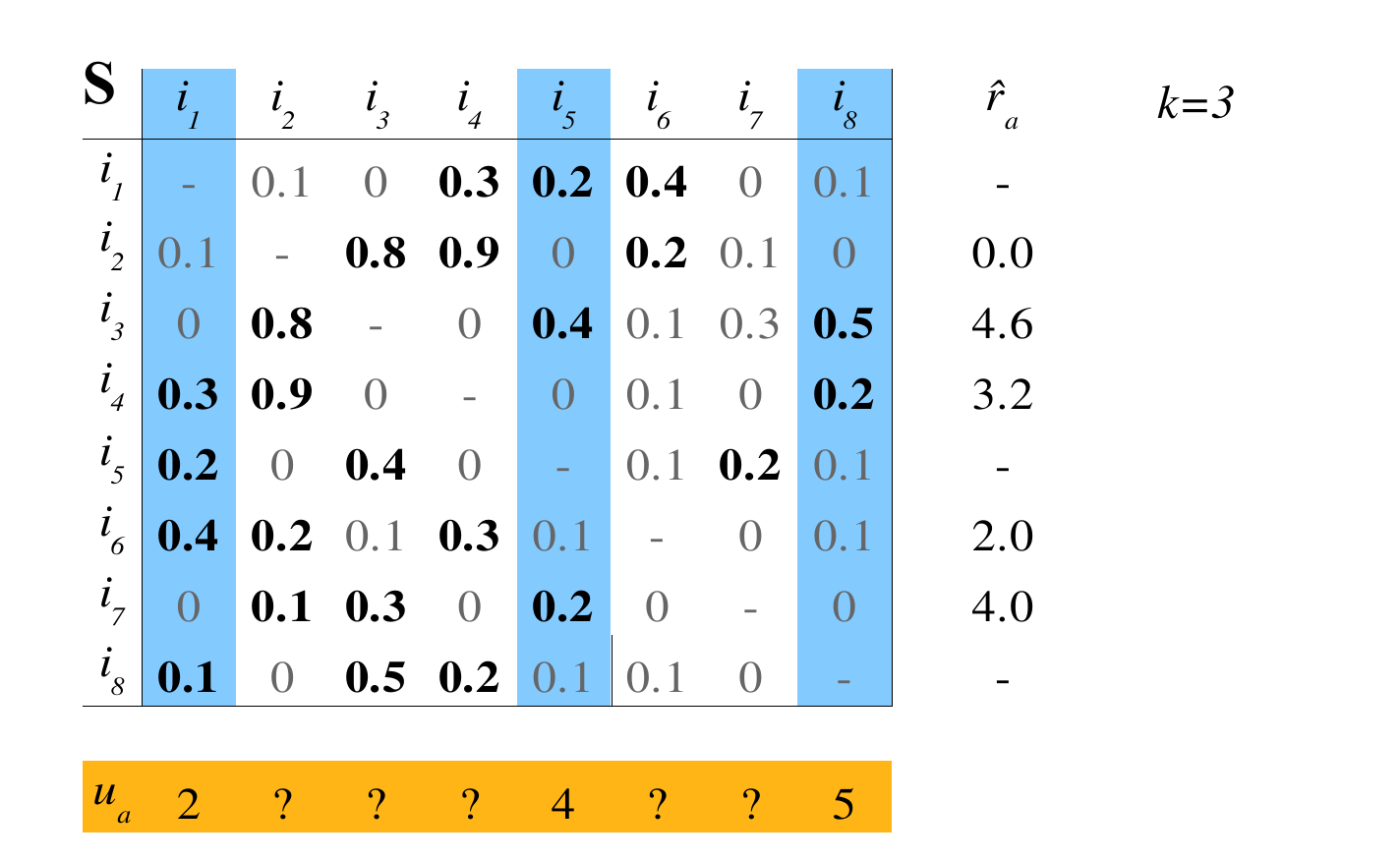}}
\caption{Item-based collaborative filtering}
\label{fig:IBCF}
\end{figure}

\begin{equation}
    \hat{r}_{ai} =  \frac{1}{\sum_{j \in \set{S}(i)\cap \{l\,;\,r_{al} \ne ?\}}{s_{ij}}}
    \sum_{j \in \set{S}(i)\cap \{l\,;\,r_{al} \ne ?\}}{s_{ij} r_{aj}}
\end{equation}

Figure~\ref{fig:IBCF} shows an example for $n=8$ items with $k=3$. For the
similarity matrix $\mat{S}$ only the $k=3$ largest entries are
stored per row (these entries are marked using bold face).
For the example we assume that we have ratings for the
active user for items
$i_1, i_5$ and $i_8$. The rows corresponding to these items are highlighted
in the item similarity matrix. We can now compute the weighted sum using
the similarities (only the reduced matrix with the $k=3$
highest ratings is used) and the user's ratings.
The result (below the matrix) shows that $i_3$ has the highest estimated
rating for the active user.

Similar to user-based recommender algorithms, user-bias can be reduced
by first normalizing the user-item rating matrix before computing the
item-to-item similarity matrix.

Item-based CF is more efficient than user-based CF since the
model (reduced similarity matrix) is relatively small ($N \times k$) and
can be fully precomputed. Item-based CF is known to
only produce slightly inferior results compared to user-based
CF and higher order models
which take the joint distribution of
sets of items into account
are possible~\citep{recommender:Deshpande:2004}.
Furthermore, item-based CF is successfully applied
in large scale recommender systems (e.g., by Amazon.com).

\subsection{User and Item-Based CF using 0-1 Data}

Less research
is available for situations where no large amount of detailed
directly elicited rating data is available.
However, this is a common situation and occurs when
users do not want to directly reveal their preferences by rating an item
(e.g., because it is to time consuming). In
this case preferences can only be inferred by analyzing usage behavior.
For
example, we can easily record in a supermarket setting what items a customer
purchases. However, we do not know why other products were not purchased.
The reason might be one of the following.
\begin{itemize}
\item The customer does not need the product right now.
\item The customer does not know about the product. Such a product
is a good candidate for recommendation.
\item The customer does not like the product. Such a product should
obviously not
be recommended.
\end{itemize}

\cite{recommender:Mild:2003} and \cite{recommender:Lee:2005}
present and evaluate
recommender algorithms for this setting.
The same reasoning is true for
recommending pages of a web site given click-stream data. Here we only have
information about which pages were viewed but not why some pages were not
viewed. This situation leads to
binary data or more exactly to
0-1 data where 1 means that we inferred that
the user has a preference for an item and 0 means that either the user does not like the item or does not know about it.
\cite{recommender:Pan:2008} call
this type of data in the context of collaborative filtering
analogous to similar situations for classifiers
\emph{one-class data} since only the 1-class is pure and contains only
positive examples. The 0-class is a mixture of positive and negative examples.

In the 0-1 case with $r_{jk} \in {0,1}$ where we define:
\begin{equation}
r_{jk}=
\begin{cases}
1& \text{user $u_j$ is known to have a preference for item $i_k$} \\
0& \text{otherwise.}
\end{cases}
\end{equation}

Two strategies to deal with one-class data is to assume all missing ratings
(zeros) are negative examples or to assume that all missing ratings are unknown.
In addition,
\cite{recommender:Pan:2008} propose strategies which represent a
trade-off between the two extreme strategies based on wighted low rank
approximations of the rating matrix and on negative example sampling which
might improve results across all recommender algorithms.

If we assume that users  typically favor only a small fraction of the items and thus most items with no rating will be indeed negative examples.
then we have no missing values and can use the approaches
described above for real valued rating data. However,
if we assume all zeroes are missing values, then this lead
to the problem that we cannot compute similarities using
Pearson correlation or Cosine similarity since the
not missing parts of the vectors only contains ones.
A similarity measure which
only focuses on matching ones
and thus prevents the problem with zeroes
is the \emph{Jaccard index}:
\begin{equation}
\mathrm{sim_{Jaccard}}(\set{X},\set{Y}) = \frac{|\set{X}\cap \set{Y}|}
{|\set{X}\cup \set{Y}|},
\end{equation}
where
$\set{X}$ and $\set{Y}$ are the sets of the items with a 1 in user profiles
$u_a$ and $u_b$, respectively.
The Jaccard index can be used between users for user-based filtering
and between items for item-based filtering as described above.


\subsection{Recommendations for 0-1 Data Based on Association Rules}
\label{sec:AR}
Recommender systems using association rules
produce recommendations based on a dependency model for items
given by a set of association
rules~\citep{Fu:2000,Mobasher:2001,Geyer-Schulz:2002,Lin:2002,Demiriz:2004}.
The binary profile matrix $\mat{R}$ is seen as a database
where each user is treated as a transaction that contains
the subset of items in $\set{I}$ with a rating of 1.
Hence transaction $k$ is defined as
$\set{T}_k = \{i_j \in \set{I} | r_{jk} = 1\}$ and
the whole transaction data base is
$\set{D} = \{\set{T}_1, \set{T}_2, \ldots, \set{T}_U\}$ where $U$ is the number of users.
To build the dependency model,
a set of association rules $\set{R}$ is mined from
$\mat{R}$. Association rules are rules of the form
$\set{X} \rightarrow \set{Y}$  where $\set{X}, \set{Y} \subseteq \set{I}$
and $\set{X} \cap \set{Y} = \emptyset$.
For the model we only use association rules with a single item in
the right-hand-side of the rule ($|\set{Y}| = 1$).
To select a set of useful association rules,
thresholds on measures of significance and interestingness are used. Two widely applied measures are:
\begin{equation*}
\mathrm{support}(\set{X} \rightarrow \set{Y}) =
\mathrm{support}(\set{X} \cup \set{Y}) =
\mathrm{Freq}(\set{X} \cup \set{Y}) / |\set{D}| = \hat{P}(E_\set{X} \cap E_\set{Y})
\end{equation*}
\begin{equation*}
\mathrm{confidence}(\set{X} \rightarrow \set{Y}) = \mathrm{support}(\set{X} \cup \set{Y}) / \mathrm{support}(\set{X}) = \hat{P}(E_\set{Y}|E_\set{X})
\end{equation*}

$\mathrm{Freq}(\set{I})$ gives the number of transactions
in the data base $\set{D}$ that contains all items in $\set{I}$.
$E_\set{I}$ is the event that the itemset $\set{I}$ is contained in a transaction.

We now require $\mathrm{support}(\set{X} \rightarrow \set{Y}) > s$ and
$\mathrm{confidence}(\set{X} \rightarrow \set{Y}) > c$
and also include a length constraint $|\set{X} \cup \set{Y}|\leq l$.
The set of rules $\set{R}$ that satisfy these constraints form the dependency
model. Although finding all association rules given thresholds
on support and confidence is a hard problem (the model grows in the
worse case exponential with the number of items), algorithms that efficiently
find all rules in most cases are
available~\citep[e.g.,][]{arules:Agrawal:1994,arules:Zaki:2000,arules:Han:2004}. Also model size can be controlled by $l$, $s$ and $c$.

To make a recommendation for an active user $u_a$ given the
set of items $\set{T}_a$ the user likes and the
set of association rules $\set{R}$ (dependency model),
the following steps are necessary:
\begin{enumerate}
\item Find all matching rules $\set{X} \rightarrow \set{Y}$ for
which $\set{X} \subseteq \set{T}_a$
in $\set{R}$.
\item Recommend $N$ unique right-hand-sides ($\set{Y}$) of the matching rules
with the highest confidence (or another measure of interestingness).
\end{enumerate}

The dependency model is very similar to
item-based CF with conditional probability-based
similarity~\citep{recommender:Deshpande:2004}. It can be fully precomputed
and rules with more than one items in the left-hand-side ($\set{X}$),
it incorporates higher order effects between more than two items.

\subsection{Other collaborative filtering methods}

Over time several other model-based approaches have been developed.
A popular simple item-based approach is the \emph{Slope One}
algorithm \citep{recommender:Lemire:2005}.
Another family of algorithms is based on latent factors approach using
matrix decomposition~\citep{recommender:Koren:2009}.
More recently, deep learning has become a very popular method for
flexible matrix completion, matrix factorization and collaborative ranking. A
comprehensive survey is presented by~\cite{recommender:Zhang:2019}.

These algorithms are outside the scope of this introductory paper.

\section{Evaluation of Recommender Algorithms}
\label{sec:evaluation}

Evaluation of recommender systems is an important topic and
reviews were presented
by \cite{recommender:Herlocker:2004}
and \cite{recommender:Gunawardana:2009}.
Typically,
given a rating matrix $\mat{R}$,
recommender algorithms are evaluated by first
partitioning the users (rows) in $\mat{R}$ into two
sets $\set{U}_\mathit{train}\, \cup\, \set{U}_\mathit{test} = \set{U}$.
The rows of $\mat{R}$ corresponding to the training users $U_\mathit{train}$
are used to learn the recommender model.
Then each user $u_a \in \set{U}_\mathit{test}$ is seen as an active user.
Before creating recommendations some items are withheld from the
profile $r_{u_a\cdot}$ and it measured
either how well the predicted rating matches the withheld value
or, for top-$N$ algorithms, if the items in the recommended
list are rated highly by the user. Finally, the evaluation measures calculated for all
test users are averaged.


To determine how to split $\set{U}$ into
$\set{U}_\mathit{train}$ and $\set{U}_\mathit{test}$ we can use
several approaches~\citep{recommender:Kohavi:1995}.
\begin{itemize}
\item {\bf Splitting:}
We can randomly assign a predefined proportion of the users
to the training set and all others to the test set.
\item {\bf Bootstrap sampling:}
We can sample from $\set{U}_\mathit{test}$ with replacement
to create the training set and then use the users not in the training set as
the test set. This procedure has the advantage that for smaller data sets
we can create larger training sets and still have users left for testing.
\item {\bf $k$-fold cross-validation:} Here we split $\set{U}$ into $k$ sets
(called folds) of approximately the same size. Then we evaluate $k$ times,
always using one fold for testing and all other folds for leaning. The $k$
results can be averaged. This approach makes sure that each user is at least
once in the test set and the averaging produces more robust results and error
estimates.
\end{itemize}

The items withheld in the test data are randomly chosen.
\cite{recommender:Breese:1998} introduced the four experimental protocols
called {\em Given 2}, {\em Given 5}, {\em Given 10} and {\em All-but-1}. For
the Given $x$ protocols for each user $x$ randomly chosen items are given to
the recommender algorithm and the remaining items are withheld for evaluation.
For All but $x$ the algorithm gets all but $x$ withheld items.

In the following we discuss the evaluation of predicted ratings
and then of top-$N$ recommendation lists.

\subsection{Evaluation of predicted ratings}

A typical way to evaluate a prediction is to compute
the deviation of the prediction from the true value.
This is the basis for the \emph{Mean
Average Error (MAE)}

\begin{equation}
\mathrm{MAE} = \frac{1}{|\set{K}|} \sum_{(i,j)\in \set{K}} |r_{ij} - \hat{r}_{ij}|,
\end{equation}
where $\set{K}$ is the set of all user-item pairings  $(i,j)$
for which we have a predicted rating $\hat{r}_{ij}$ and a known rating
$r_{ij}$ which was not used to learn the recommendation model.

Another popular measure is the \emph{Root Mean Square Error (RMSE)}.
\begin{equation}
\mathrm{RMSE} = \sqrt{\frac{\sum_{(i,j)\in \set{K}} (r_{ij} - \hat{r}_{ij})^2}{|\set{K}|}}
\end{equation}

RMSE penalizes larger errors stronger than MAE and thus is suitable for
situations where small prediction errors are not very important.

\subsection{Evaluation Top-$N$ recommendations}

The items in the predicted top-$N$ lists and the
withheld items  liked by the user (typically determined by a simple threshold
on the actual rating) for all test users $\set{U}_{test}$
can be aggregated into a so called {\em confusion matrix} depicted in
table~\ref{tab_confusion} (see \cite{recommender:Kohavi:1998})
which corresponds
exactly to the outcomes of a classical statistical experiment.
The confusion matrix shows how many of the items recommended in
the top-$N$ lists (column predicted positive; $d+b$)
were withheld items and thus correct recommendations
(cell $d$) and how many where potentially incorrect (cell $b$).
The matrix also shows
how many of the not recommended items
(column predicted negative; $a+c$)
should have actually been recommended since they represent withheld items (cell $c$).

\begin{table}[tbp]
\caption{2x2 confusion matrix
\label{tab_confusion}
}
\center
\begin{tabular}{|c||c|c|}
\hline
{\bf actual / predicted} & {\bf negative}  & {\bf positive} \\
\hline
\hline
{\bf negative} & $a$ & $b$ \\
\hline
{\bf positive} & $c$ & $d$ \\
\hline
\end{tabular}
\end{table}

From the confusion matrix several performance measures can be derived.
For the data mining task of a recommender system
the performance of an algorithm depends on its ability
to learn significant patterns in the data set.
Performance measures used to evaluate these algorithms
have their root in machine learning.
A commonly used measure is
{\em accuracy,} the fraction of correct
recommendations to total possible recommendations.

\begin{equation}
\mathit{Accuracy} = \frac{\mathit{correct\ recommendations}}{\mathit{total\ possible\ recommendations}}
=  \frac{a+d}{a+b+c+d}
\label{accur}
\end{equation}

A common error measure is the {\em mean absolute error} ({\em MAE}, also called {\em mean absolute deviation} or {\em MAD}).
\begin{equation}
\mathit{MAE} = \frac{1}{N}\sum_{i=1}^N{|\epsilon_i|} = \frac{b+c}{a+b+c+d},
\label{mae}
\end{equation}
where $N = a+b+c+d$ is the total number of items
which can be recommended and $|\epsilon_i|$ is the
absolute error of each item.
Since we deal with 0-1 data, $|\epsilon_i|$ can only be zero (in cells $a$ and $d$ in the confusion matrix) or one (in cells $b$ and $c$).
For evaluation recommender algorithms for rating data, the root mean square
error is often used. For 0-1 data it reduces to the square
root of MAE.

Recommender systems help to find items of interest from the set of all
available items. This can be seen as a retrieval task known from
information retrieval. Therefore, standard information retrieval
performance measures are frequently used to evaluate recommender performance.
{\em Precision} and {\em recall} are the best known measures used in
information retrieval \citep{recommender:Salton:1983,recommender:Rijsbergen:1979}.

\begin{equation}
\mathit{Precision} = \frac{\mathit{correctly\ recommended\ items}}{\mathit{total\ recommended\ items}}
= \frac{d}{b+d}
\label{prec}
\end{equation}

\begin{equation}
\mathit{Recall} = \frac{\mathit{correctly\ recommended\ items}}{\mathit{total\ useful\ recommendations}}
= \frac{d}{c+d}
\label{rec}
\end{equation}


Often the number of {\em total\ useful\ recommendations} needed for recall is
unknown since the whole collection would have to be inspected. However, instead
of the actual {\em total\ useful\ recommendations} often the total number of
known useful recommendations is used.  Precision and recall are
conflicting properties, high precision means low recall and vice versa. To find
an optimal trade-off between precision and recall a single-valued measure like
the {\em E-measure} \citep{recommender:Rijsbergen:1979} can be used.
The parameter $\alpha$ controls the
trade-off between precision and recall.
\begin{equation}
\mbox{\textit{E-measure}} = \frac{1}{\alpha(1/\mathit{Precision}) + (1-\alpha)(1/\mathit{Recall})}
\label{e-measure}
\end{equation}

A popular single-valued measure is the {\em F-measure.} It is defined as the
harmonic mean of precision and recall.
\begin{equation}
\mbox{\textit{F-measure}} = \frac{2\, \mathit{Precision}\, \mathit{Recall}}{\mathit{Precision} + \mathit{Recall}} =
\frac{2}{1/\mathit{Precision} + 1/\mathit{Recall}}
\label{f-measure}
\end{equation}
It is
a special case of the E-measure with $\alpha=.5$ which places the same weight
on both, precision and recall.  In the recommender evaluation literature the
F-measure is often referred to as the measure {\em F1.}

Another method used in the literature to compare two classifiers at different
parameter settings is the {\em Receiver Operating Characteristic (ROC)}.
The method was developed for signal detection and goes back to the Swets model
\citep{recommender:Rijsbergen:1979}. The ROC-curve is a plot of the system's
{\em probability of detection} (also called $\mathit{sensitivity}$ or true
positive rate TPR which is equivalent to recall as defined in
formula \ref{rec}) by the {\em probability of false alarm}
(also called false positive rate FPR or
$1-\mathit{specificity}$, where
$\mathit{specificity} = \frac{a}{a+b}$) with
regard to model parameters.  A possible way to compare the efficiency of two
systems is by comparing the size of the area under the ROC-curve, where a
bigger area indicates better performance.

\section{Recommenderlab Infrastructure}
\label{sec:infrastructure}
\begin{figure}
    \centerline{\includegraphics[width=1\textwidth, trim={0 23cm 0 0}, clip]{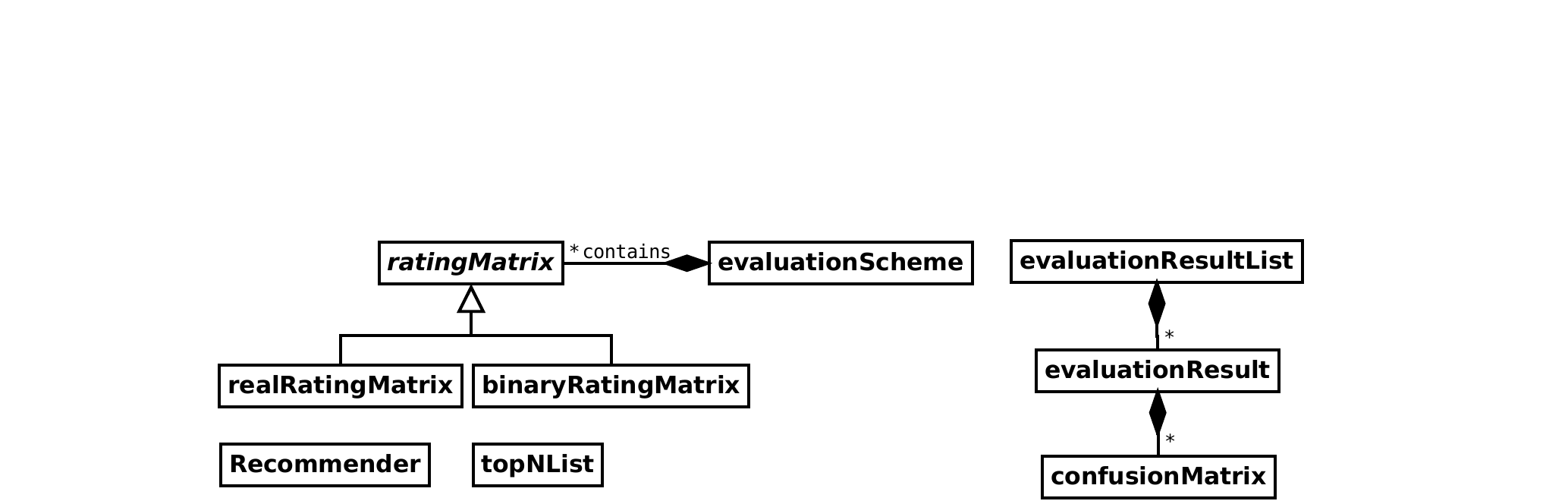}}
\caption{UML class diagram for
package~\pkg{recommenderlab}~\citep{misc:Fowler:2004}.}
\label{fig:class_diagram}
\end{figure}

\pkg{recommenderlab} is implemented using formal classes in
the \proglang{S4} class system.
Figure~\ref{fig:class_diagram} shows the main classes and their relationships.

The package uses the abstract \class{ratingMatrix}
to provide a common interface for rating data. \class{ratingMatrix}
implements many methods typically available for matrix-like
objects. For example, \func{dim}, \func{dimnames},
\func{colCounts}, \func{rowCounts},
\func{colMeans}, \func{rowMeans},
\func{colSums} and \func{rowSums}.
Additionally \func{sample} can be used to sample from users (rows) and
\func{image} produces an image plot.

For \class{ratingMatrix} we provide two concrete
implementations \class{realRatingMatrix} and
\class{binaryRatingMatrix} to represent different types of rating
matrices $\mat{R}$.
\class{realRatingMatrix} implements a rating matrix with real valued
ratings stored in sparse format defined in package \pkg{Matrix}.
Sparse matrices in \pkg{Matrix} typically do not store 0s explicitly,
however for \class{realRatingMatrix} we use these sparse matrices
such that instead of 0s, NAs are not explicitly stored.

\class{binaryRatingMatrix} implements a 0-1 rating matrix using the
implementation of \class{itemMatrix} defined in package~\pkg{arules}.
\class{itemMatrix} stores only the ones and internally uses a sparse
representation from package \pkg{Matrix}.  With this class structure
\pkg{recommenderlab} can be easily extended to other forms of rating matrices
with different concepts for efficient storage in the future.

Class \class{Recommender} implements the data structure to store recommendation
models. The creator method

\begin{center}
\code{Recommender(data, method, parameter = NULL)}
\end{center}

takes data as a \class{ratingMatrix},
a method name and some optional parameters for the method and
returns a \class{Recommender} object. Once we have a recommender object,
we can predict top-$N$ recommendations for active users using

\begin{center}
\code{predict(object, newdata, n=10, type=c("topNList", "ratings", "ratingMatrix"), ...)}.

\end{center}

Predict can return either top-$N$ lists (default setting) or predicted ratings.
\code{object} is the recommender
object, \code{newdata} is the data for the active users.
For top-$N$ lists \code{n} is
the maximal number of recommended items in each list and \func{predict} will
return an objects of class \class{topNList} which contains
one top-$N$ list for each active user. For \code{"ratings"}
and \code{"ratingMatrix"}, \code{n} is ignored
and an object of \class{realRatingMatrix} is returned. Each row
contains the predicted ratings for one active user. The difference is,
that for \code{"ratings"}, the items for which a
rating exists in \code{newdata} have a \code{NA} instead of a
predicted/actual ratings.

The actual implementations for the recommendation algorithms are
managed using the registry mechanism provided by package \pkg{registry}.
The registry called \code{recommenderRegistry}
and stores recommendation method names and a short description.
Generally, the registry mechanism is hidden from the user and the
creator function \func{Recommender} uses it in the background to map
a recommender method name to its implementation. However, the
registry can be directly queried by

\begin{center}
\code{recommenderRegistry\$get\_entries()}
\end{center}

and new recommender algorithms can be added by the user.
We will give and example for this feature
in the examples section of this paper.

To evaluate recommender algorithms package~\pkg{recommenderlab}
provides the infrastructure to create and maintain evaluation schemes
stored as an object of class \class{evaluationScheme} from rating data.
The creator function

\begin{center}
\code{evaluationScheme(data, method="split", train=0.9, k=10, given=3)}
\end{center}

creates the evaluation scheme from a data set using a method
(e.g., simple split, bootstrap sampling, $k$-fold cross validation).
Testing is perfomed by withholding items (parameter \code{given}).
\cite{recommender:Breese:1998} introduced the
four experimental witholding protocols called \emph{Given 2, Given 5, Given 10} and \emph{All-but-1.}
During testing, the \emph{Given $x$} protocol presents the algorithm with
only $x$ randomly chosen items for the test user, and the algorithm
is evaluated by how well it is able to predict the withheld items.
For \emph{All-but-$x$}, a generalization of \emph{All-but-1},
the algorithm sees all but
$x$ withheld ratings for the test user.
\code{given} controls x in the evaluations scheme.
Positive integers result in a Given $x$ protocol, while negative values
produce a All-but-$x$ protocol.

The function \func{evaluate} is then used to
evaluate several recommender algorithms using an evaluation scheme resulting
in a evaluation result list (class~\class{evaluationResultList}) with
one entry (class~\class{evaluationResult}) per algorithm.
Each object of \class{evaluationResult}
contains one or several object of \class{confusionMatrix} depending on the
number of evaluations specified in the \class{evaluationScheme} (e.g., $k$
for $k$-fold cross validation).
With this infrastructure several recommender algorithms can be compared
on a data set with a single line of code.

In the following, we will illustrate the usage of \pkg{recommenderlab}
with several examples.
\section{Examples}
\label{sec:examples}
This fist few example shows how to manage data in recommender lab and then
we create and evaluate recommenders.
First, we load the package.

\begin{Schunk}
\begin{Sinput}
R> library("recommenderlab")
\end{Sinput}
\end{Schunk}

\subsection{Coercion to and from rating matrices}

For this example we create a small
artificial data set as a matrix.

\begin{Schunk}
\begin{Sinput}
R> m <- matrix(sample(c(as.numeric(0:5), NA), 50,
+     replace=TRUE, prob=c(rep(.4/6,6),.6)), ncol=10,
+     dimnames=list(user=paste("u", 1:5, sep=''),
+ 	item=paste("i", 1:10, sep='')))
R> m
\end{Sinput}
\begin{Soutput}
    item
user i1 i2 i3 i4 i5 i6 i7 i8 i9 i10
  u1 NA  2  3  5 NA  5 NA  4 NA  NA
  u2  2 NA NA NA NA NA NA NA  2   3
  u3  2 NA NA NA NA  1 NA NA NA  NA
  u4  2  2  1 NA NA  5 NA  0  2  NA
  u5  5 NA NA NA NA NA NA  5 NA   4
\end{Soutput}
\end{Schunk}

With coercion, the matrix can be easily converted into
a realRatingMatrix object which stores the data in sparse
format (only non-NA values are stored explicitly; NA values are
represented by a dot).

\begin{Schunk}
\begin{Sinput}
R> r <- as(m, "realRatingMatrix")
R> r
\end{Sinput}
\begin{Soutput}
5 x 10 rating matrix of class ???realRatingMatrix??? with 19 ratings.
\end{Soutput}
\begin{Sinput}
R> getRatingMatrix(r)
\end{Sinput}
\begin{Soutput}
5 x 10 sparse Matrix of class "dgCMatrix"
                               
u1 . 2 3 5 . 5 .  4.000e+00 . .
u2 2 . . . . . . .          2 3
u3 2 . . . . 1 . .          . .
u4 2 2 1 . . 5 . 2.225e-308 2 .
u5 5 . . . . . .  5.000e+00 . 4
\end{Soutput}
\end{Schunk}

The realRatingMatrix can be coerced back into a matrix which is
identical to the original matrix.
\begin{Schunk}
\begin{Sinput}
R> identical(as(r, "matrix"),m)
\end{Sinput}
\begin{Soutput}
[1] TRUE
\end{Soutput}
\end{Schunk}

It can also be coerced into a list
of users with their ratings for closer inspection
or into a data.frame with user/item/rating tuples.
\begin{Schunk}
\begin{Sinput}
R> as(r, "list")
\end{Sinput}
\begin{Soutput}
$`0`
i2 i3 i4 i6 i8 
 2  3  5  5  4 

$`1`
 i1  i9 i10 
  2   2   3 

$`2`
i1 i6 
 2  1 

$`3`
        i1         i2         i3         i6         i8         i9 
 2.000e+00  2.000e+00  1.000e+00  5.000e+00 2.225e-308  2.000e+00 

$`4`
 i1  i8 i10 
  5   5   4 
\end{Soutput}
\begin{Sinput}
R> head(as(r, "data.frame"))
\end{Sinput}
\begin{Soutput}
   user item rating
5    u1   i2      2
7    u1   i3      3
9    u1   i4      5
10   u1   i6      5
13   u1   i8      4
1    u2   i1      2
\end{Soutput}
\end{Schunk}

The data.frame version is especially suited for writing rating data to a
file (e.g., by \func{write.csv}).
Coercion from data.frame (user/item/rating tuples)
and list into a sparse rating matrix is also provided.
This way, external rating data can easily be imported into \pkg{recommenderlab}.

\subsection{Normalization}
An important operation for rating matrices is to normalize the entries
to, e.g., centering to remove rating bias by subtracting the row mean from
all ratings in the row. This is can be easily done using
\code{normalize()}.

\begin{Schunk}
\begin{Sinput}
R> r_m <- normalize(r)
R> r_m
\end{Sinput}
\begin{Soutput}
5 x 10 rating matrix of class ???realRatingMatrix??? with 19 ratings.
Normalized using center on rows.
\end{Soutput}
\begin{Sinput}
R> getRatingMatrix(r_m)
\end{Sinput}
\begin{Soutput}
5 x 10 sparse Matrix of class "dgCMatrix"
                                                                        
u1  .           -1.800e+00 -0.8 1.2 .  1.2 .  0.2000  .           .     
u2  -3.333e-01  .           .   .   .  .   .  .       -3.333e-01  0.6667
u3   5.000e-01  .           .   .   . -0.5 .  .       .           .     
u4  2.225e-308  2.225e-308 -1.0 .   .  3.0 . -2.0000  2.225e-308  .     
u5   3.333e-01  .           .   .   .  .   .  0.3333  .          -0.6667
\end{Soutput}
\end{Schunk}

Normalization can be reversed using \code{denormalize()}.
\begin{Schunk}
\begin{Sinput}
R> denormalize(r_m)
\end{Sinput}
\begin{Soutput}
5 x 10 rating matrix of class ???realRatingMatrix??? with 19 ratings.
\end{Soutput}
\end{Schunk}

Small portions of rating matrices can be visually inspected using
\code{image()}.

\begin{Schunk}
\begin{Sinput}
R> image(r, main = "Raw Ratings")
R> image(r_m, main = "Normalized Ratings")
\end{Sinput}
\end{Schunk}

\begin{figure}
\begin{minipage}[b]{.48\linewidth}
\centerline{\includegraphics[width=\linewidth]{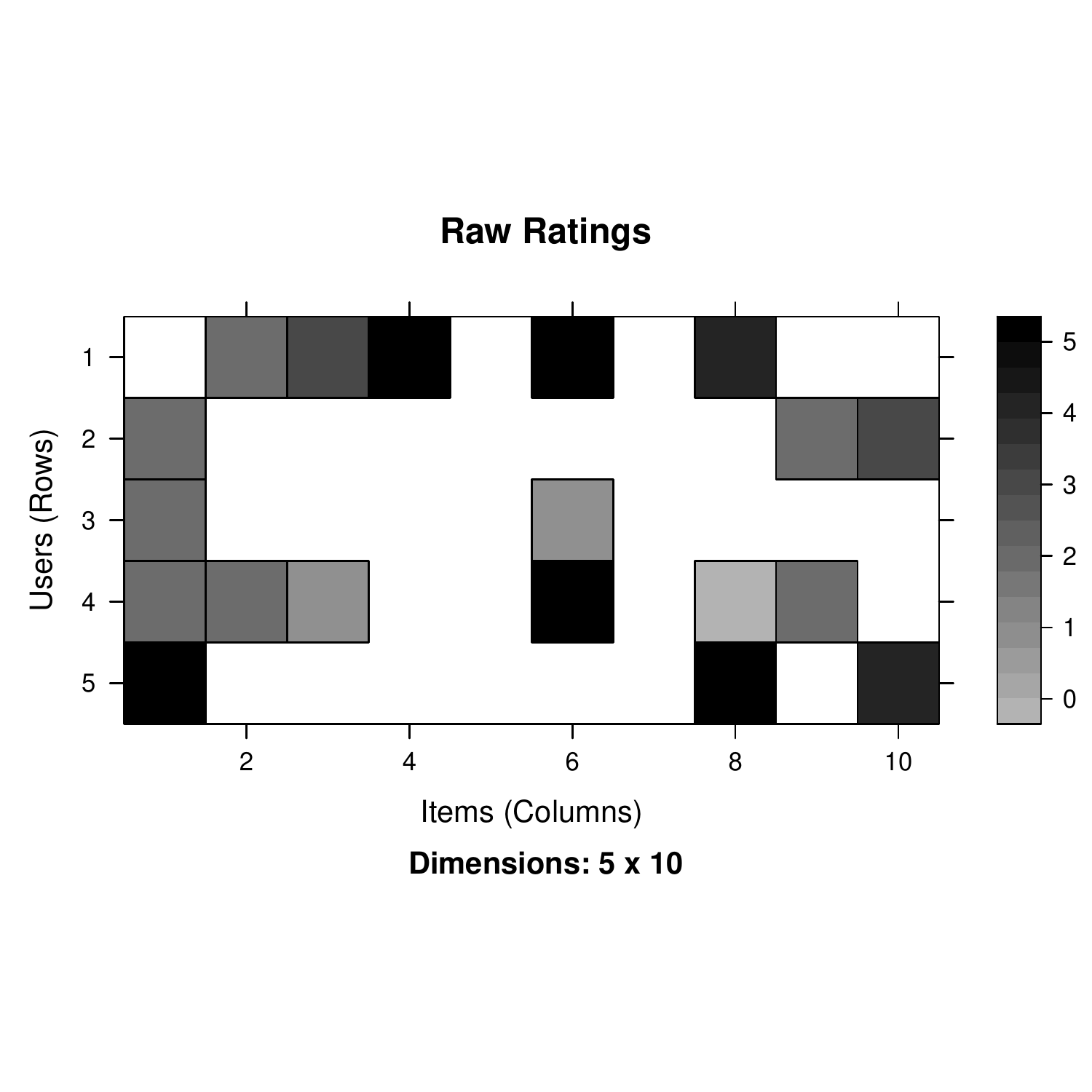}}
\end{minipage}
\begin{minipage}[b]{.48\linewidth}
\centerline{\includegraphics[width=\linewidth]{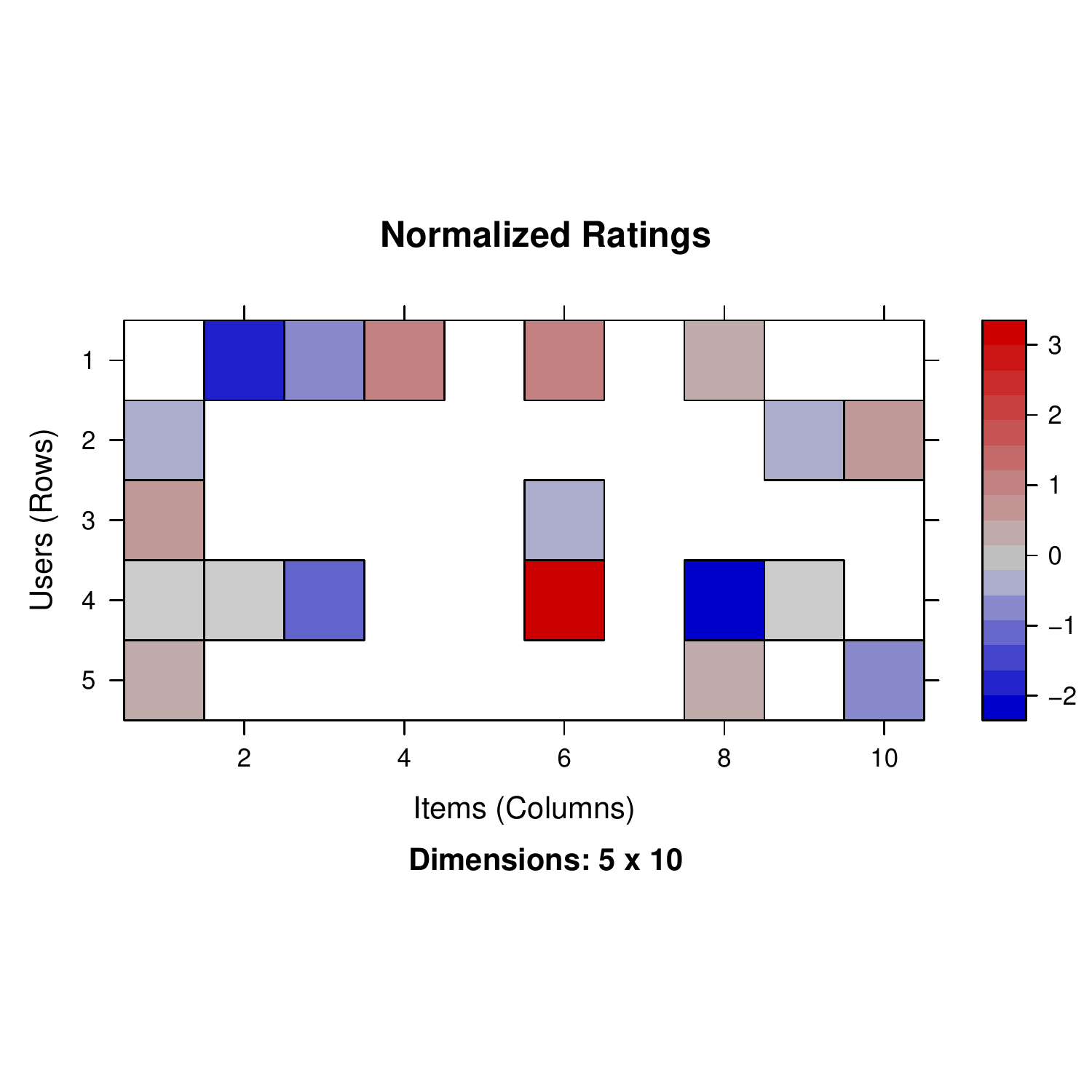}}
\end{minipage}
\caption{Image plot
the artificial rating data before and after normalization.}
\label{fig:image1}
\end{figure}

Figure~\ref{fig:image1} shows the resulting plots.

\subsection{Binarization of data}

A matrix with real valued ratings can be transformed into a 0-1 matrix
with \code{binarize()} and
a user specified threshold (\code{min\_ratings})
on the raw or normalized ratings. In the following only
items with a rating of 4 or higher will become a positive
rating in the new binary rating matrix.
\begin{Schunk}
\begin{Sinput}
R> r_b <- binarize(r, minRating=4)
R> r_b
\end{Sinput}
\begin{Soutput}
5 x 10 rating matrix of class ???binaryRatingMatrix??? with 7 ratings.
\end{Soutput}
\begin{Sinput}
R> as(r_b, "matrix")
\end{Sinput}
\begin{Soutput}
      i1    i2    i3    i4    i5    i6    i7    i8    i9   i10
u1 FALSE FALSE FALSE  TRUE FALSE  TRUE FALSE  TRUE FALSE FALSE
u2 FALSE FALSE FALSE FALSE FALSE FALSE FALSE FALSE FALSE FALSE
u3 FALSE FALSE FALSE FALSE FALSE FALSE FALSE FALSE FALSE FALSE
u4 FALSE FALSE FALSE FALSE FALSE  TRUE FALSE FALSE FALSE FALSE
u5  TRUE FALSE FALSE FALSE FALSE FALSE FALSE  TRUE FALSE  TRUE
\end{Soutput}
\end{Schunk}

\subsection{Inspection of data set properties}
We will use the data set Jester5k for the rest of this section.  This data set
comes with \pkg{recommenderlab} and contains a sample of 5000 users from the
anonymous ratings data from the Jester Online Joke Recommender System collected
between April 1999 and May 2003~\citep{recommender:Goldberg:2001}. The
data set contains ratings for 100 jokes on a scale from $-10$ to $+10$. All
users in the data set have rated 36 or more jokes.

\begin{Schunk}
\begin{Sinput}
R> data(Jester5k)
R> Jester5k
\end{Sinput}
\begin{Soutput}
5000 x 100 rating matrix of class ???realRatingMatrix??? with 363209 ratings.
\end{Soutput}
\end{Schunk}

Jester5k contains
363209~ratings. For the following examples we use
only a subset of the data containing a sample of 1000 users (we set the
random number generator seed for reproducibility).
For random sampling \func{sample} is provided for rating matrices.

\begin{Schunk}
\begin{Sinput}
R> set.seed(1234)
R> r <- sample(Jester5k, 1000)
R> r
\end{Sinput}
\begin{Soutput}
1000 x 100 rating matrix of class ???realRatingMatrix??? with 74323 ratings.
\end{Soutput}
\end{Schunk}

This subset still contains 74323~ratings.
Next, we inspect the ratings for the first user. We can select
an individual user with the extraction operator.

\begin{Schunk}
\begin{Sinput}
R> rowCounts(r[1,])
\end{Sinput}
\begin{Soutput}
u20648 
    74 
\end{Soutput}
\begin{Sinput}
R> as(r[1,], "list")
\end{Sinput}
\begin{Soutput}
$`0`
   j1    j2    j3    j4    j5    j6    j7    j8    j9   j10   j11   j12 
-2.86  1.75 -4.03 -5.78  2.23 -5.44 -3.40  8.74 -4.51  3.74  0.15 -0.39 
  j13   j14   j15   j16   j17   j18   j19   j20   j21   j22   j23   j24 
-1.94  5.97 -7.77  1.26  2.18 -2.14  1.17 -8.64 -1.36  1.21  4.95 -9.81 
  j25   j26   j27   j28   j29   j30   j31   j32   j33   j34   j35   j36 
-3.35  3.01  2.33  1.36  9.08 -7.72 -9.42  0.97 -5.83 -0.83  6.36  3.54 
  j37   j38   j39   j40   j41   j42   j43   j44   j45   j46   j47   j48 
-3.64  0.87  2.23  3.54 -7.96 -1.41  2.62 -8.45 -0.29 -9.76 -4.47  3.11 
  j49   j50   j51   j52   j53   j54   j55   j56   j57   j58   j59   j60 
 6.26  4.95 -9.17 -8.01  5.49 -5.97  1.70  5.00  4.13 -2.18  0.49  2.77 
  j61   j62   j63   j64   j65   j66   j67   j68   j69   j70   j72   j81 
-3.01  8.35 -2.77 -3.25  5.39  5.49 -1.31 -3.74  2.96  0.15  4.81  5.15 
  j85   j92 
 2.23 -8.40 
\end{Soutput}
\begin{Sinput}
R> rowMeans(r[1,])
\end{Sinput}
\begin{Soutput}
 u20648 
-0.4232 
\end{Soutput}
\end{Schunk}

The user has rated 74 jokes, the list shows
the ratings and the user's rating average is -0.423243243243243 .

Next, we look at several distributions to understand the data better.
\code{getRatings()} extracts a vector with all
non-missing
ratings from a rating matrix.

\begin{Schunk}
\begin{Sinput}
R> hist(getRatings(r), breaks=100)
\end{Sinput}
\end{Schunk}
\begin{figure}
\centerline{\includegraphics[width=.5\linewidth]{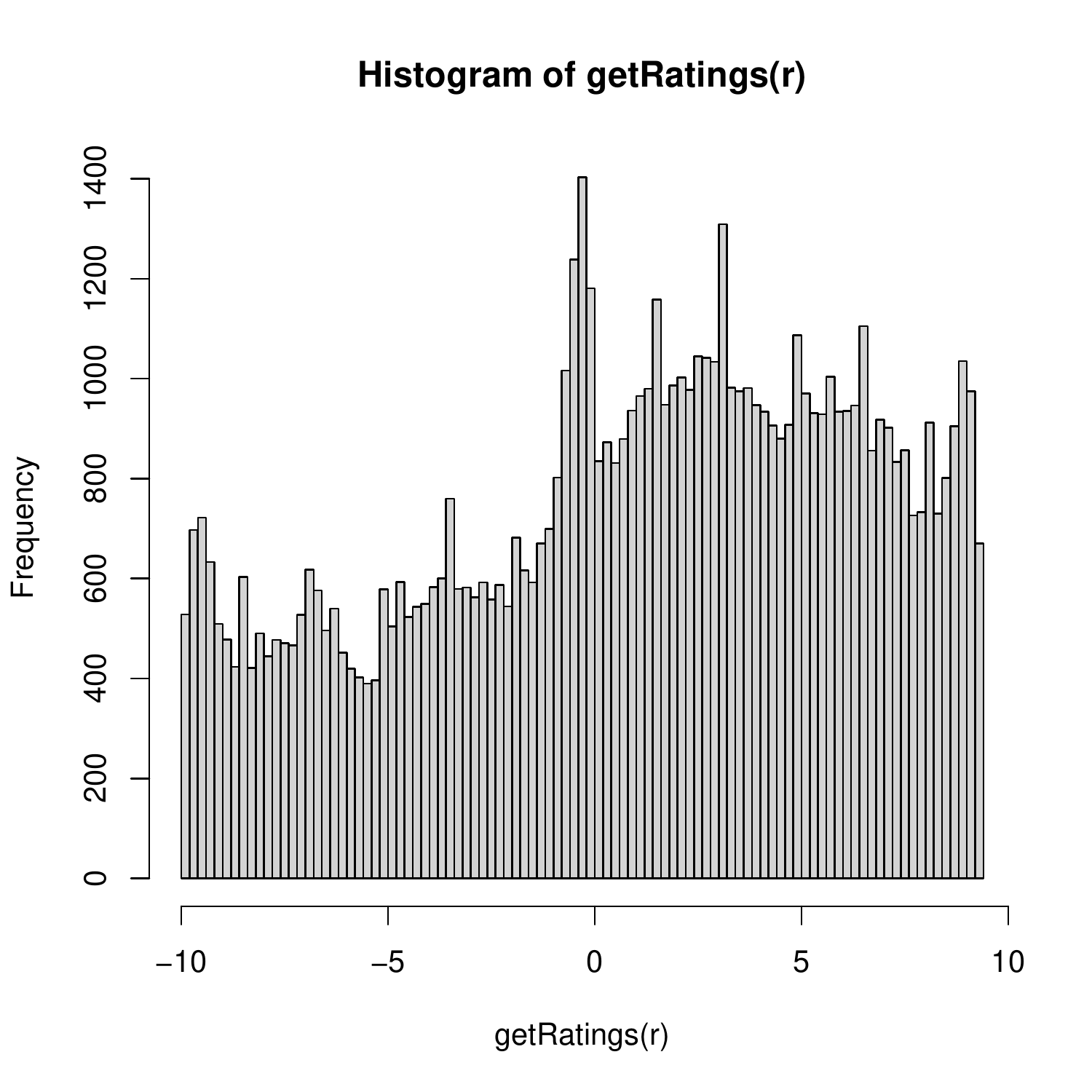}}
\caption{Raw rating distribution for as sample of Jester.}
\label{fig:hist1}
\end{figure}

In the histogram in Figure~\ref{fig:hist1}
shoes an interesting distribution
where all negative values occur with a almost identical frequency and
the positive ratings more frequent with a steady decline towards the
rating 10. Since this distribution can be the result of users with strong
rating bias, we look next at the rating distribution after normalization.

\begin{Schunk}
\begin{Sinput}
R> hist(getRatings(normalize(r)), breaks=100)
\end{Sinput}
\end{Schunk}
\begin{Schunk}
\begin{Sinput}
R> hist(getRatings(normalize(r, method="Z-score")), breaks=100)
\end{Sinput}
\end{Schunk}

\begin{figure}
\begin{minipage}[b]{.48\linewidth}
\centerline{\includegraphics[width=\linewidth]{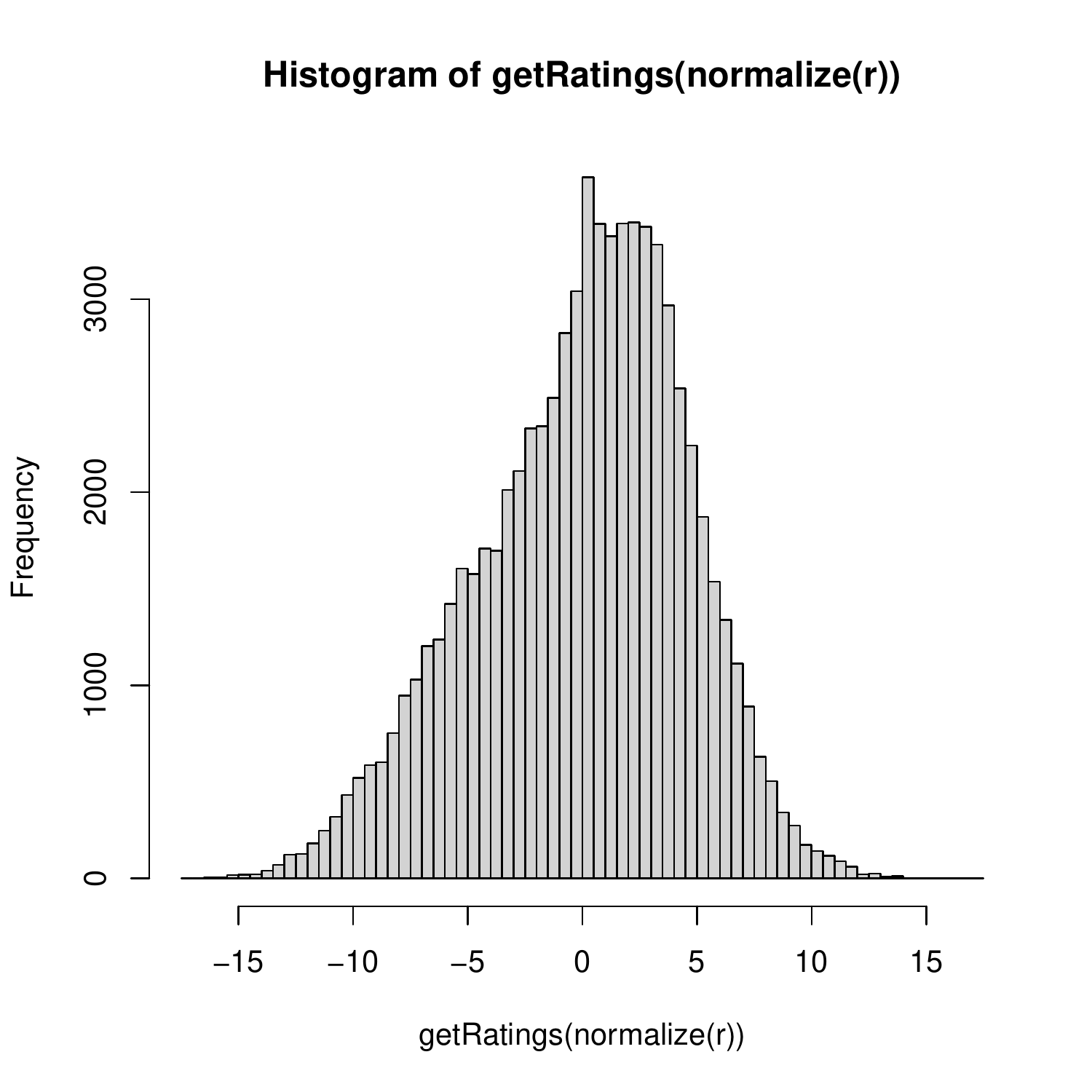}}
\end{minipage}
\begin{minipage}[b]{.48\linewidth}
\centerline{\includegraphics[width=\linewidth]{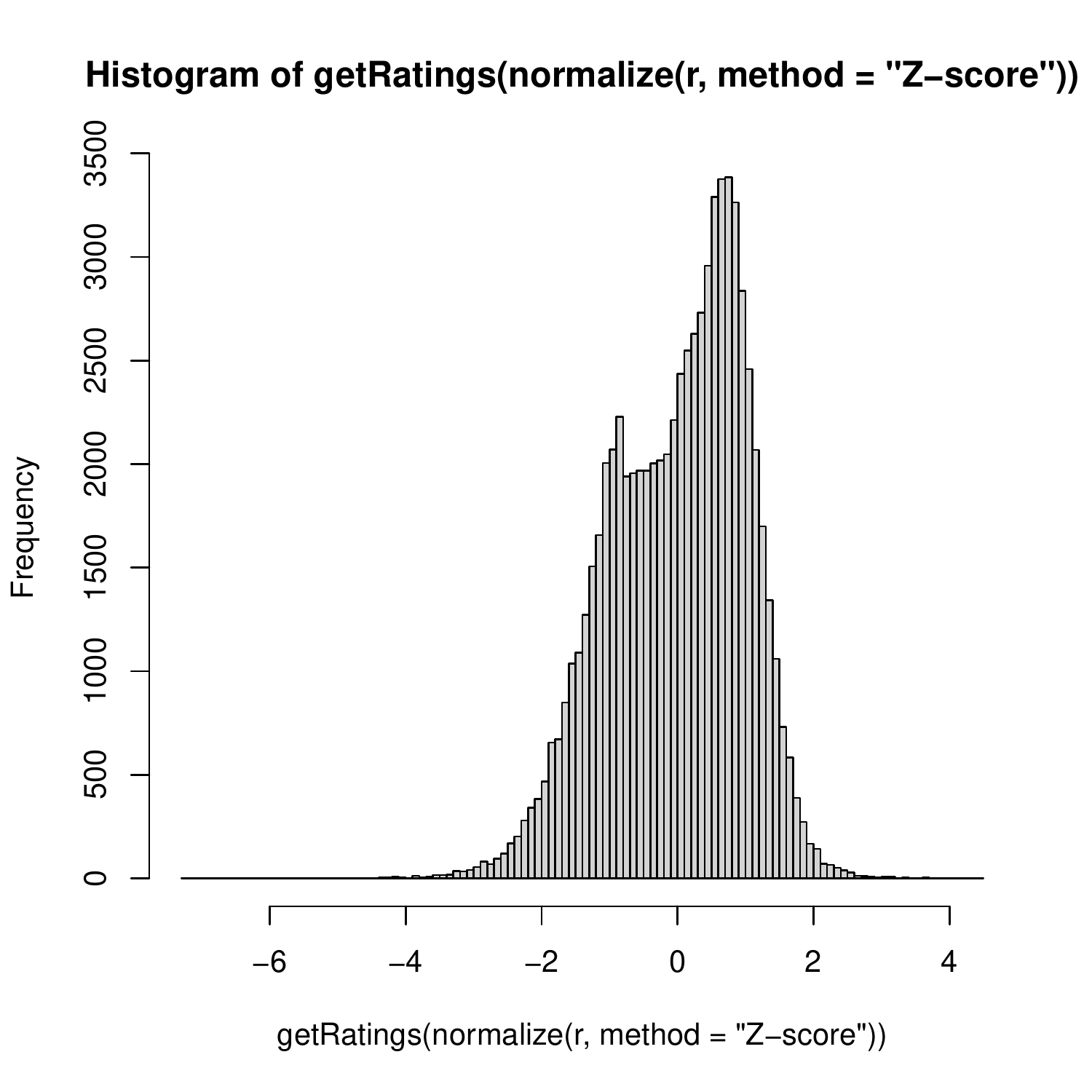}}
\end{minipage}
\caption{Histogram of normalized ratings using row centering (left) and
Z-score normalization (right).}
\label{fig:hist2}
\end{figure}

Figure~\ref{fig:hist2} shows that the distribution of ratings
ins closer to a normal distribution after row centering and
Z-score normalization additionally reduces the variance to a
range of roughly $-3$ to $+3$ standard deviations. It is interesting to see
that there is a pronounced peak of ratings between zero and two.

Finally, we look at how many jokes each user has rated and what the
mean rating for each Joke is.

\begin{Schunk}
\begin{Sinput}
R> hist(rowCounts(r), breaks=50)
\end{Sinput}
\end{Schunk}

\begin{Schunk}
\begin{Sinput}
R> hist(colMeans(r), breaks=20)
\end{Sinput}
\end{Schunk}

\begin{figure}
\begin{minipage}[b]{.48\linewidth}
\centerline{\includegraphics[width=\linewidth]{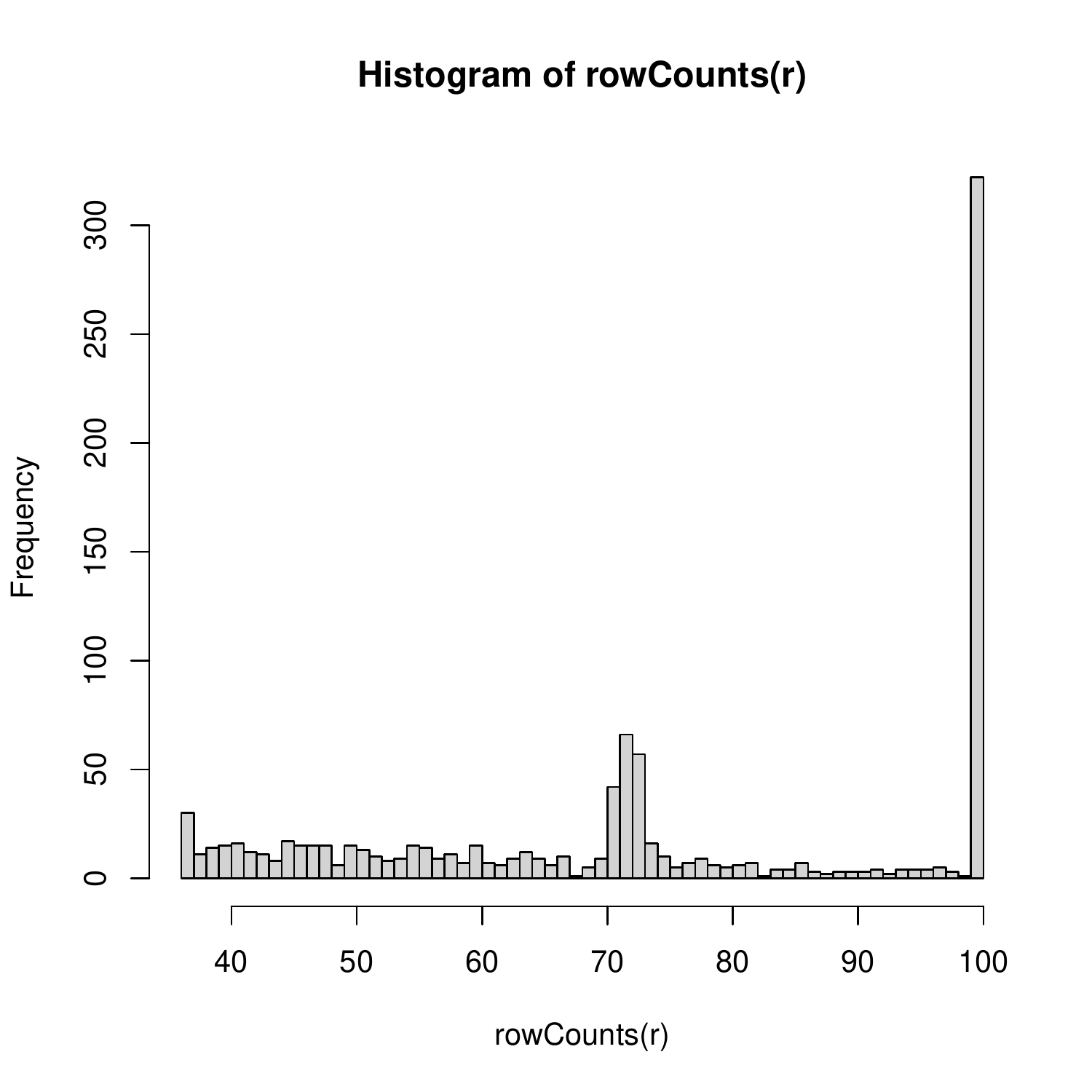}}
\end{minipage}
\begin{minipage}[b]{.48\linewidth}
\centerline{\includegraphics[width=\linewidth]{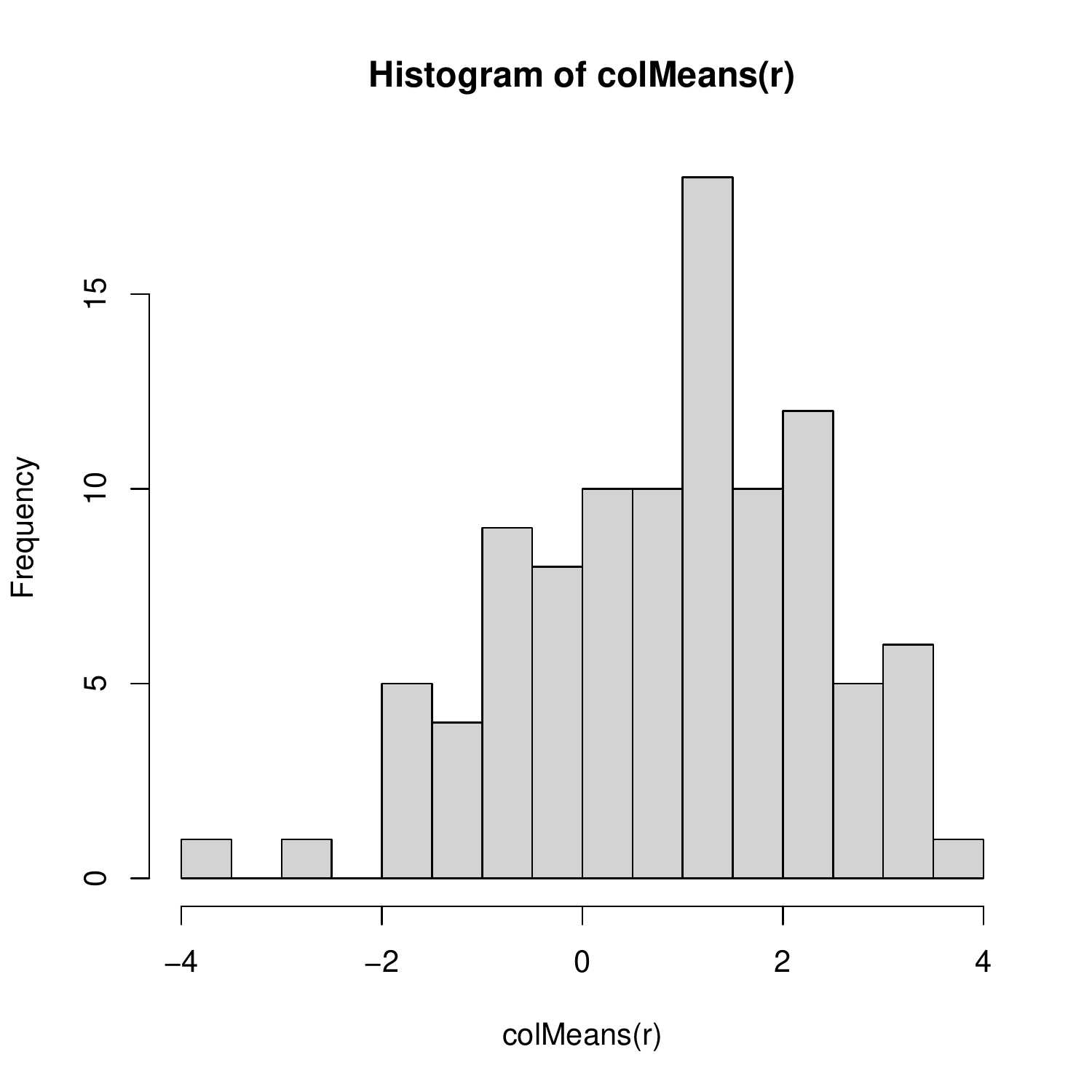}}
\end{minipage}
\caption{Distribution of the number of rated items per user (left) and
of the average ratings per joke (right).}
\label{fig:hist3}
\end{figure}

Figure~\ref{fig:hist3} shows that there are unusually many
users with ratings around 70 and users who have rated all jokes.
The average ratings per joke look closer to a normal distribution with
a mean above 0.

\subsection{Creating a recommender}

A recommender is created using the creator function \func{Recommender}.
Available recommendation methods are stored in a registry.
The registry can be queried. Here we are only interested in
methods for real-valued rating data.

\begin{Schunk}
\begin{Sinput}
R> recommenderRegistry$get_entries(dataType = "realRatingMatrix")
\end{Sinput}
\begin{Soutput}
$HYBRID_realRatingMatrix
Recommender method: HYBRID for realRatingMatrix
Description: Hybrid recommender that aggegates several recommendation strategies using weighted averages.
Reference: NA
Parameters:
  recommenders weights aggregation_type
1         NULL    NULL            "sum"

$ALS_realRatingMatrix
Recommender method: ALS for realRatingMatrix
Description: Recommender for explicit ratings based on latent factors, calculated by alternating least squares algorithm.
Reference: Yunhong Zhou, Dennis Wilkinson, Robert Schreiber, Rong Pan (2008). Large-Scale Parallel Collaborative Filtering for the Netflix Prize, 4th Int'l Conf. Algorithmic Aspects in Information and Management, LNCS 5034.
Parameters:
  normalize lambda n_factors n_iterations min_item_nr seed
1      NULL    0.1        10           10           1 NULL

$ALS_implicit_realRatingMatrix
Recommender method: ALS_implicit for realRatingMatrix
Description: Recommender for implicit data based on latent factors, calculated by alternating least squares algorithm.
Reference: Yifan Hu, Yehuda Koren, Chris Volinsky (2008). Collaborative Filtering for Implicit Feedback Datasets, ICDM '08 Proceedings of the 2008 Eighth IEEE International Conference on Data Mining, pages 263-272.
Parameters:
  lambda alpha n_factors n_iterations min_item_nr seed
1    0.1    10        10           10           1 NULL

$IBCF_realRatingMatrix
Recommender method: IBCF for realRatingMatrix
Description: Recommender based on item-based collaborative filtering.
Reference: NA
Parameters:
   k   method normalize normalize_sim_matrix alpha na_as_zero
1 30 "Cosine"  "center"                FALSE   0.5      FALSE

$LIBMF_realRatingMatrix
Recommender method: LIBMF for realRatingMatrix
Description: Matrix factorization with LIBMF via package recosystem (https://cran.r-project.org/web/packages/recosystem/vignettes/introduction.html).
Reference: NA
Parameters:
  dim costp_l2 costq_l2 nthread
1  10     0.01     0.01       1

$POPULAR_realRatingMatrix
Recommender method: POPULAR for realRatingMatrix
Description: Recommender based on item popularity.
Reference: NA
Parameters:
  normalize
1  "center"
                                                     aggregationRatings
1 new("standardGeneric", .Data = function (x, na.rm = FALSE, dims = 1, 
                                                  aggregationPopularity
1 new("standardGeneric", .Data = function (x, na.rm = FALSE, dims = 1, 

$RANDOM_realRatingMatrix
Recommender method: RANDOM for realRatingMatrix
Description: Produce random recommendations (real ratings).
Reference: NA
Parameters: None

$RERECOMMEND_realRatingMatrix
Recommender method: RERECOMMEND for realRatingMatrix
Description: Re-recommends highly rated items (real ratings).
Reference: NA
Parameters:
  randomize minRating
1         1        NA

$SVD_realRatingMatrix
Recommender method: SVD for realRatingMatrix
Description: Recommender based on SVD approximation with column-mean imputation.
Reference: NA
Parameters:
   k maxiter normalize
1 10     100  "center"

$SVDF_realRatingMatrix
Recommender method: SVDF for realRatingMatrix
Description: Recommender based on Funk SVD with gradient descend (https://sifter.org/~simon/journal/20061211.html).
Reference: NA
Parameters:
   k gamma lambda min_epochs max_epochs min_improvement normalize verbose
1 10 0.015  0.001         50        200        0.000001  "center"   FALSE

$UBCF_realRatingMatrix
Recommender method: UBCF for realRatingMatrix
Description: Recommender based on user-based collaborative filtering.
Reference: NA
Parameters:
    method nn sample weighted normalize min_matching_items
1 "cosine" 25  FALSE     TRUE  "center"                  0
  min_predictive_items
1                    0
\end{Soutput}
\end{Schunk}

Next, we create a recommender which generates recommendations
solely on the popularity of items (the number of users who have the item in
their profile). We create a recommender
from the first 1000 users in the Jester5k data set.

\begin{Schunk}
\begin{Sinput}
R> r <- Recommender(Jester5k[1:1000], method = "POPULAR")
R> r
\end{Sinput}
\begin{Soutput}
Recommender of type ???POPULAR??? for ???realRatingMatrix??? 
learned using 1000 users.
\end{Soutput}
\end{Schunk}

The model can be obtained from a recommender using \func{getModel}.
\begin{Schunk}
\begin{Sinput}
R> names(getModel(r))
\end{Sinput}
\begin{Soutput}
[1] "topN"                  "ratings"               "normalize"            
[4] "aggregationRatings"    "aggregationPopularity" "verbose"              
\end{Soutput}
\begin{Sinput}
R> getModel(r)$topN
\end{Sinput}
\begin{Soutput}
Recommendations as ???topNList??? with n = 100 for 1 users. 
\end{Soutput}
\end{Schunk}

In this case the model has a
top-$N$ list to store the popularity order and further
elements (average ratings, if it used normalization and the
used aggregation function).

Recommendations are generated by \func{predict}
(consistent with its use for other types of models in
\proglang{R}). The result
are recommendations in the form of an object of class~\class{TopNList}.
Here we create top-5 recommendation lists for two
users who were not used to learn the model.

\begin{Schunk}
\begin{Sinput}
R> recom <- predict(r, Jester5k[1001:1002], n=5)
R> recom
\end{Sinput}
\begin{Soutput}
Recommendations as ???topNList??? with n = 5 for 2 users. 
\end{Soutput}
\end{Schunk}

The result contains two ordered top-$N$ recommendation lists,
one for each user. The recommended items can be inspected as a list.
\begin{Schunk}
\begin{Sinput}
R> as(recom, "list")
\end{Sinput}
\begin{Soutput}
$u15553
[1] "j89" "j72" "j93" "j76" "j87"

$u7886
[1] "j89" "j72" "j93" "j76" "j1" 
\end{Soutput}
\end{Schunk}

Since the top-$N$ lists are ordered, we can extract sublists of
the best items in the top-$N$. For example, we can get the best 3
recommendations for each list using \func{bestN}.
\begin{Schunk}
\begin{Sinput}
R> recom3 <- bestN(recom, n = 3)
R> recom3
\end{Sinput}
\begin{Soutput}
Recommendations as ???topNList??? with n = 3 for 2 users. 
\end{Soutput}
\begin{Sinput}
R> as(recom3, "list")
\end{Sinput}
\begin{Soutput}
$u15553
[1] "j89" "j72" "j93"

$u7886
[1] "j89" "j72" "j93"
\end{Soutput}
\end{Schunk}

Many recommender algorithms can also predict ratings. This is also
implemented using \func{predict} with the parameter \code{type}
set to \code{"ratings"}.

\begin{Schunk}
\begin{Sinput}
R> recom <- predict(r, Jester5k[1001:1002], type="ratings")
R> recom
\end{Sinput}
\begin{Soutput}
2 x 100 rating matrix of class ???realRatingMatrix??? with 72 ratings.
\end{Soutput}
\begin{Sinput}
R> as(recom, "matrix")[,1:10]
\end{Sinput}
\begin{Soutput}
          j1    j2    j3    j4 j5 j6 j7 j8    j9 j10
u15553    NA    NA    NA    NA NA NA NA NA    NA  NA
u7886  4.005 2.918 2.944 0.998 NA NA NA NA 2.319  NA
\end{Soutput}
\end{Schunk}

Predicted ratings are returned as an object of \class{realRatingMatrix}.
The prediction contains \code{NA} for the items rated by the active users.
In the example we show the predicted ratings for the first 10 items for
the two active users.

Alternatively, we can also request the complete rating matrix which includes
the original ratings by the user.

\begin{Schunk}
\begin{Sinput}
R> recom <- predict(r, Jester5k[1001:1002], type="ratingMatrix")
R> recom
\end{Sinput}
\begin{Soutput}
2 x 100 rating matrix of class ???realRatingMatrix??? with 200 ratings.
\end{Soutput}
\begin{Sinput}
R> as(recom, "matrix")[,1:10]
\end{Sinput}
\begin{Soutput}
          j1    j2    j3    j4    j5    j6    j7    j8    j9   j10
u15553 6.756 5.669 5.695 3.749 5.616 6.989 4.683 4.802 5.070 6.790
u7886  4.005 2.918 2.944 0.998 2.865 4.238 1.932 2.050 2.319 4.039
\end{Soutput}
\end{Schunk}

\subsection{Evaluation of predicted ratings}
Next, we will look at the evaluation of recommender algorithms.
\pkg{recommenderlab} implements several
standard evaluation methods for recommender systems.
Evaluation starts with creating an evaluation scheme
that determines what and how data is used for training and testing.
Here we create an evaluation scheme which splits the first 1000 users
in Jester5k into a training set (90\%) and a test set (10\%). For
the test set 15 items will be given to the recommender algorithm
and the other items will be held out
for computing the error.

\begin{Schunk}
\begin{Sinput}
R> e <- evaluationScheme(Jester5k[1:1000], method="split", train=0.9,
+     given=15, goodRating=5)
R> e
\end{Sinput}
\begin{Soutput}
Evaluation scheme with 15 items given
Method: ???split??? with 1 run(s).
Training set proportion: 0.900
Good ratings: >=5.000000
Data set: 1000 x 100 rating matrix of class ???realRatingMatrix??? with 74164 ratings.
\end{Soutput}
\end{Schunk}

We create two recommenders (user-based and item-based collaborative filtering)
using the training data.

\begin{Schunk}
\begin{Sinput}
R> r1 <- Recommender(getData(e, "train"), "UBCF")
R> r1
\end{Sinput}
\begin{Soutput}
Recommender of type ???UBCF??? for ???realRatingMatrix??? 
learned using 900 users.
\end{Soutput}
\begin{Sinput}
R> r2 <- Recommender(getData(e, "train"), "IBCF")
R> r2
\end{Sinput}
\begin{Soutput}
Recommender of type ???IBCF??? for ???realRatingMatrix??? 
learned using 900 users.
\end{Soutput}
\end{Schunk}

Next, we compute predicted ratings for the known part of the test data (15 items
for each user) using
the two algorithms.

\begin{Schunk}
\begin{Sinput}
R> p1 <- predict(r1, getData(e, "known"), type="ratings")
R> p1
\end{Sinput}
\begin{Soutput}
100 x 100 rating matrix of class ???realRatingMatrix??? with 8357 ratings.
\end{Soutput}
\begin{Sinput}
R> p2 <- predict(r2, getData(e, "known"), type="ratings")
R> p2
\end{Sinput}
\begin{Soutput}
100 x 100 rating matrix of class ???realRatingMatrix??? with 8417 ratings.
\end{Soutput}
\end{Schunk}

Finally, we can calculate the error between the prediction and the
unknown part of the test data.

\begin{Schunk}
\begin{Sinput}
R> error <- rbind(
+   UBCF = calcPredictionAccuracy(p1, getData(e, "unknown")),
+   IBCF = calcPredictionAccuracy(p2, getData(e, "unknown"))
+ )
R> error
\end{Sinput}
\begin{Soutput}
      RMSE   MSE   MAE
UBCF 4.571 20.89 3.585
IBCF 4.538 20.60 3.440
\end{Soutput}
\end{Schunk}

In this example user-based collaborative filtering produces a smaller
prediction error.

\subsection{Evaluation of a top-$N$ recommender algorithm}

For this example we create a $4$-fold cross validation scheme
with the the Given-3 protocol, i.e.,
for the test users all but three randomly selected items are withheld
for evaluation.

\begin{Schunk}
\begin{Sinput}
R> scheme <- evaluationScheme(Jester5k[1:1000], method="cross", k=4, given=3,
+     goodRating=5)
R> scheme
\end{Sinput}
\begin{Soutput}
Evaluation scheme with 3 items given
Method: ???cross-validation??? with 4 run(s).
Good ratings: >=5.000000
Data set: 1000 x 100 rating matrix of class ???realRatingMatrix??? with 74164 ratings.
\end{Soutput}
\end{Schunk}

Next we use the created evaluation scheme to evaluate the recommender
method popular.
We evaluate top-1, top-3, top-5, top-10, top-15 and top-20 recommendation lists.

\begin{Schunk}
\begin{Sinput}
R> results <- evaluate(scheme, method="POPULAR", type = "topNList",
+   n=c(1,3,5,10,15,20))
\end{Sinput}
\begin{Soutput}
POPULAR run fold/sample [model time/prediction time]
	 1  [0.008sec/0.173sec] 
	 2  [0.006sec/0.157sec] 
	 3  [0.006sec/0.153sec] 
	 4  [0.007sec/0.192sec] 
\end{Soutput}
\begin{Sinput}
R> results
\end{Sinput}
\begin{Soutput}
Evaluation results for 4 folds/samples using method ???POPULAR???.
\end{Soutput}
\end{Schunk}

The result is an object of class~\class{EvaluationResult} which
contains several confusion matrices. \func{getConfusionMatrix}
will return the confusion matrices for the 4 runs
(we used 4-fold cross evaluation) as a list.
In the following we look at the first element of the list which
represents the first of the 4 runs.

\begin{Schunk}
\begin{Sinput}
R> getConfusionMatrix(results)[[1]]
\end{Sinput}
\begin{Soutput}
        TP     FP    FN    TN  N precision  recall     TPR     FPR  n
[1,] 0.464  0.536 19.25 76.75 97    0.4640 0.03458 0.03458 0.00678  1
[2,] 1.372  1.628 18.34 75.66 97    0.4573 0.08912 0.08912 0.01988  3
[3,] 2.228  2.772 17.49 74.51 97    0.4456 0.13867 0.13867 0.03379  5
[4,] 4.352  5.648 15.36 71.64 97    0.4352 0.28256 0.28256 0.06890 10
[5,] 6.272  8.728 13.44 68.56 97    0.4181 0.40248 0.40248 0.10828 15
[6,] 7.600 12.400 12.12 64.88 97    0.3800 0.45762 0.45762 0.15522 20
\end{Soutput}
\end{Schunk}

For the first run we have 6 confusion matrices represented by rows, one for each
of the six different top-$N$ lists we used for evaluation.
$n$ is the number of recommendations per list. TP, FP, FN and TN are the
entries for
true positives, false positives, false negatives and true negatives in
the confusion matrix. The remaining columns contain precomputed performance
measures. The average for all runs can be obtained
from the evaluation results directly using \func{avg}.

\begin{Schunk}
\begin{Sinput}
R> avg(results)
\end{Sinput}
\begin{Soutput}
        TP     FP    FN    TN  N precision  recall     TPR      FPR  n
[1,] 0.453  0.547 18.12 77.88 97    0.4530 0.03428 0.03428 0.006685  1
[2,] 1.299  1.701 17.28 76.72 97    0.4330 0.09435 0.09435 0.020704  3
[3,] 2.119  2.881 16.46 75.54 97    0.4238 0.14437 0.14437 0.034955  5
[4,] 4.127  5.873 14.45 72.55 97    0.4127 0.27452 0.27452 0.071829 10
[5,] 5.922  9.078 12.65 69.35 97    0.3948 0.38732 0.38732 0.111440 15
[6,] 7.347 12.653 11.23 65.77 97    0.3673 0.46375 0.46375 0.155834 20
\end{Soutput}
\end{Schunk}

Evaluation results can be plotted using \func{plot}. The default
plot is the ROC curve which plots the true positive rate (TPR) against the
false positive rate (FPR).
\begin{Schunk}
\begin{Sinput}
R> plot(results, annotate=TRUE)
\end{Sinput}
\end{Schunk}
\begin{figure}
\centerline{\includegraphics[scale=1]{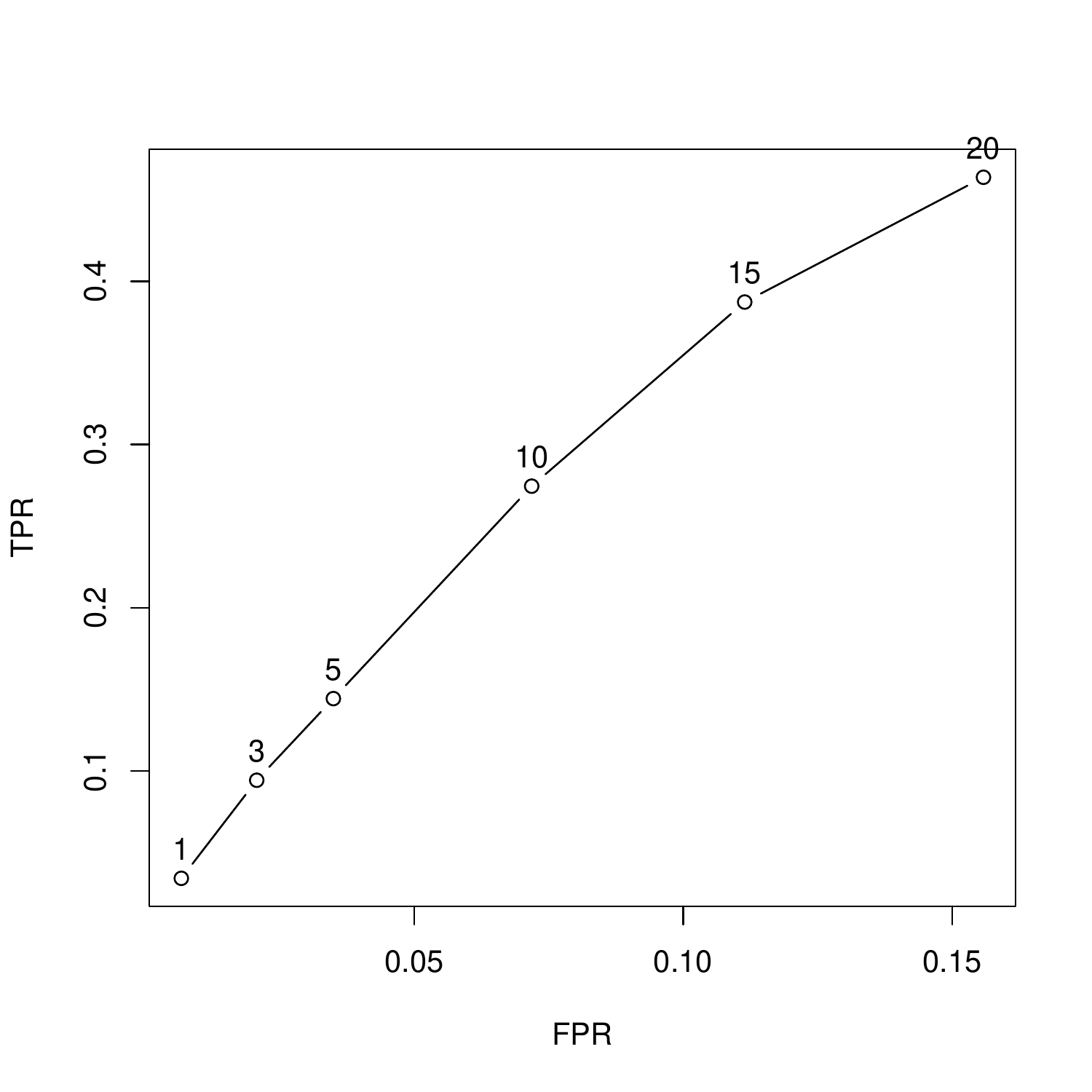}}
\caption{ROC curve for recommender method POPULAR.}
\label{fig:roc1}
\end{figure}

\begin{figure}
\centerline{\includegraphics[scale=1]{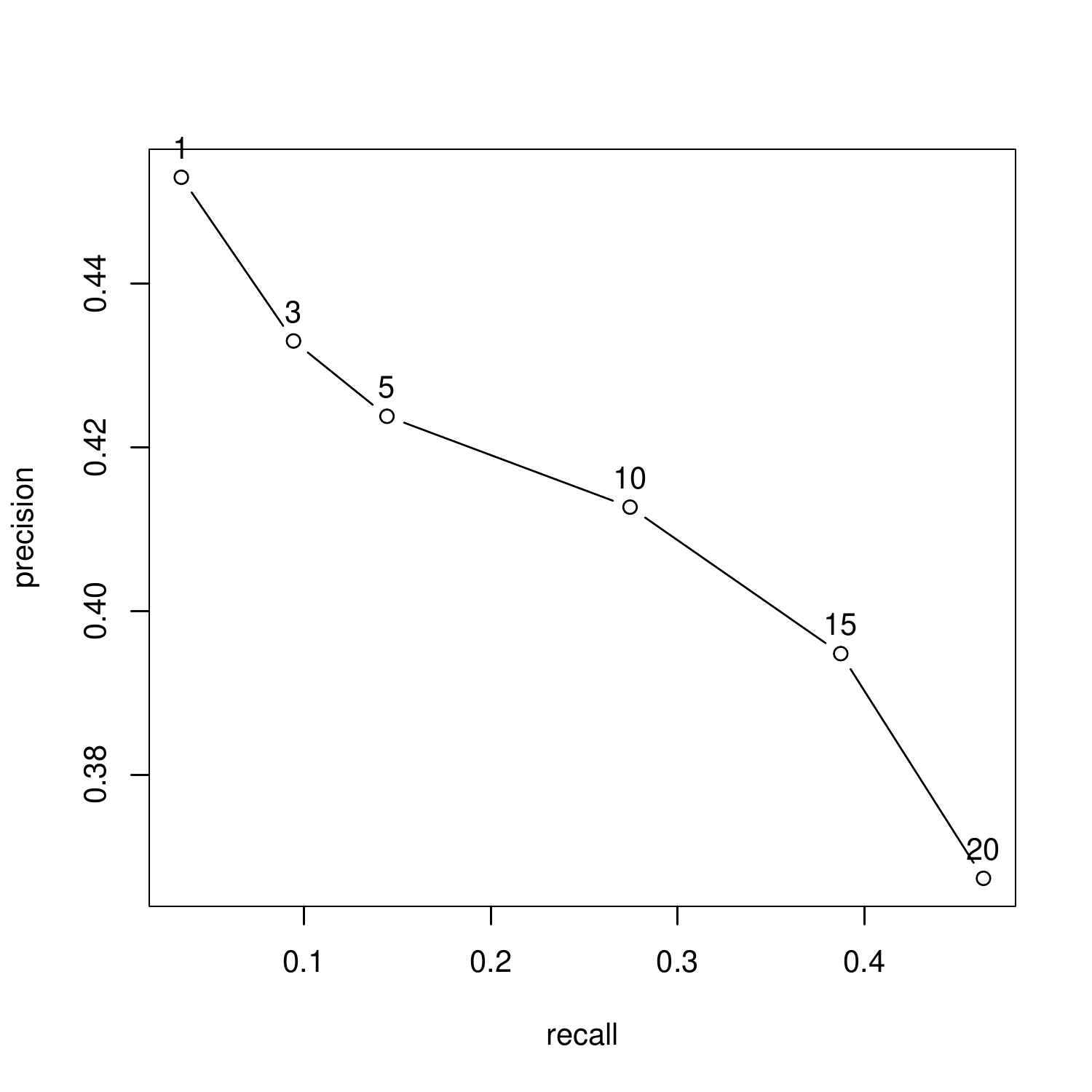}}
\caption{Precision-recall plot for method POPULAR.}
\label{fig:precrec1}
\end{figure}

For the plot where we annotated the curve with the size of the top-$N$ list
is shown in Figure~\ref{fig:roc1}.
By using \code{"prec/rec"} as the second argument, a precision-recall plot
is produced (see Figure~\ref{fig:precrec1}).

\begin{Schunk}
\begin{Sinput}
R> plot(results, "prec/rec", annotate=TRUE)
\end{Sinput}
\end{Schunk}

\subsection{Comparing recommender algorithms}
\subsubsection{Comparing top-$N$ recommendations}

The comparison of
several recommender algorithms is one of the main functions of
\pkg{recommenderlab}. For comparison also \func{evaluate} is used.
The only change is to use \func{evaluate} with
a list of algorithms together with their parameters instead of
a single method name. In the following we use the
evaluation scheme created above to compare the five
recommender algorithms: random items, popular items,
user-based CF, item-based CF, and SVD approximation.
Note that when running the following code, the CF based algorithms
are very slow.

For the evaluation we use a ``all-but-5'' scheme. This is indicated by
a negative number for \code{given}.

\begin{Schunk}
\begin{Sinput}
R> set.seed(2016)
R> scheme <- evaluationScheme(Jester5k[1:1000], method="split", train = .9,
+   k=1, given=-5, goodRating=5)
R> scheme
\end{Sinput}
\begin{Soutput}
Evaluation scheme using all-but-5 items
Method: ???split??? with 1 run(s).
Training set proportion: 0.900
Good ratings: >=5.000000
Data set: 1000 x 100 rating matrix of class ???realRatingMatrix??? with 74164 ratings.
\end{Soutput}
\begin{Sinput}
R> algorithms <- list(
+   "random items" = list(name="RANDOM", param=NULL),
+   "popular items" = list(name="POPULAR", param=NULL),
+   "user-based CF" = list(name="UBCF", param=list(nn=50)),
+   "item-based CF" = list(name="IBCF", param=list(k=50)),
+   "SVD approximation" = list(name="SVD", param=list(k = 50))
+ )
R> ## run algorithms
R> results <- evaluate(scheme, algorithms, type = "topNList",
+   n=c(1, 3, 5, 10, 15, 20))
\end{Sinput}
\begin{Soutput}
RANDOM run fold/sample [model time/prediction time]
	 1  [0.001sec/0.011sec] 
POPULAR run fold/sample [model time/prediction time]
	 1  [0.006sec/0.059sec] 
UBCF run fold/sample [model time/prediction time]
	 1  [0.01sec/0.17sec] 
IBCF run fold/sample [model time/prediction time]
	 1  [0.038sec/0.017sec] 
SVD run fold/sample [model time/prediction time]
	 1  [0.094sec/0.012sec] 
\end{Soutput}
\end{Schunk}

The result is an object of class~\class{evaluationResultList} for the
five recommender algorithms.
\begin{Schunk}
\begin{Sinput}
R> results
\end{Sinput}
\begin{Soutput}
List of evaluation results for 5 recommenders:

$`random items`
Evaluation results for 1 folds/samples using method ???RANDOM???.

$`popular items`
Evaluation results for 1 folds/samples using method ???POPULAR???.

$`user-based CF`
Evaluation results for 1 folds/samples using method ???UBCF???.

$`item-based CF`
Evaluation results for 1 folds/samples using method ???IBCF???.

$`SVD approximation`
Evaluation results for 1 folds/samples using method ???SVD???.
\end{Soutput}
\end{Schunk}

Individual results can be accessed by list subsetting using an index or
the name specified when calling \func{evaluate}.

\begin{Schunk}
\begin{Sinput}
R> names(results)
\end{Sinput}
\begin{Soutput}
[1] "random items"      "popular items"     "user-based CF"    
[4] "item-based CF"     "SVD approximation"
\end{Soutput}
\begin{Sinput}
R> results[["user-based CF"]]
\end{Sinput}
\begin{Soutput}
Evaluation results for 1 folds/samples using method ???UBCF???.
\end{Soutput}
\end{Schunk}

\begin{figure}
\centerline{\includegraphics[scale=1]{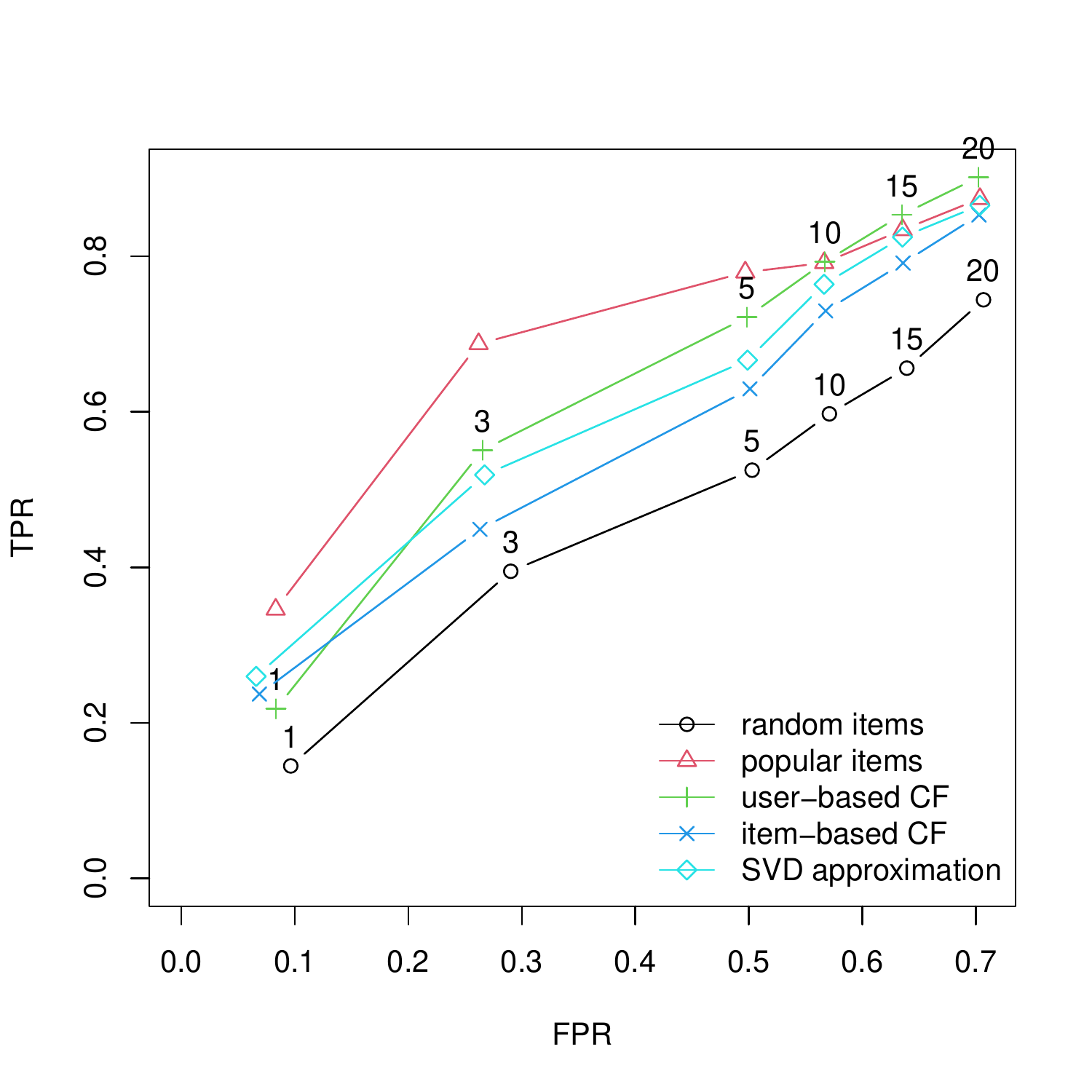}}
\caption{Comparison of ROC curves for several recommender methods for the
given-3 evaluation scheme.}
\label{fig:roc2}
\end{figure}

\begin{figure}
\centerline{\includegraphics[scale=1]{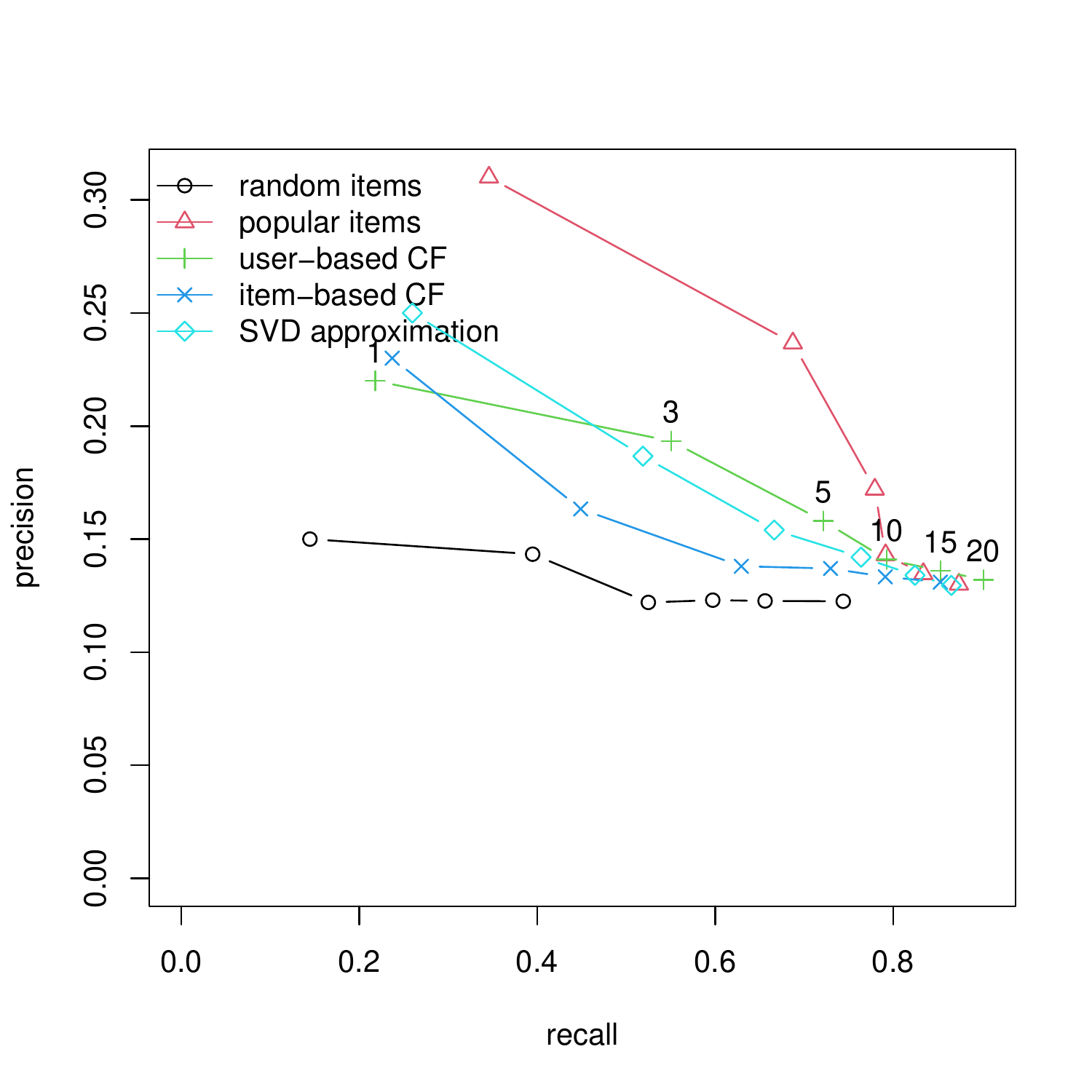}}
\caption{Comparison of precision-recall curves
for several recommender methods for the
    given-3 evaluation scheme.}
\label{fig:precrec2}
\end{figure}

Again \func{plot} can be used to create ROC and precision-recall plots
(see Figures~\ref{fig:roc2} and \ref{fig:precrec2}). Plot accepts most
of the usual graphical parameters like pch, type, lty, etc. In addition
annotate can be used to annotate the points on selected curves with the
list length.

\begin{Schunk}
\begin{Sinput}
R> plot(results, annotate=c(1,3), legend="bottomright")
\end{Sinput}
\end{Schunk}

\begin{Schunk}
\begin{Sinput}
R> plot(results, "prec/rec", annotate=3, legend="topleft")
\end{Sinput}
\end{Schunk}

For this data set and the given evaluation scheme
popular items and the user-based CF methods clearly
outperform the other methods. In Figure~\ref{fig:roc2} we see that they
dominate (almost completely) the other method since for each length of top-$N$ list they
provide a better combination of TPR and FPR.

\subsubsection{Comparing ratings}

Next, we evaluate not top-$N$ recommendations, but how well the algorithms
can predict ratings.

\begin{Schunk}
\begin{Sinput}
R> ## run algorithms
R> results <- evaluate(scheme, algorithms, type = "ratings")
\end{Sinput}
\begin{Soutput}
RANDOM run fold/sample [model time/prediction time]
	 1  [0.002sec/0.003sec] 
POPULAR run fold/sample [model time/prediction time]
	 1  [0.007sec/0.004sec] 
UBCF run fold/sample [model time/prediction time]
	 1  [0.006sec/0.146sec] 
IBCF run fold/sample [model time/prediction time]
	 1  [0.042sec/0.01sec] 
SVD run fold/sample [model time/prediction time]
	 1  [0.097sec/0.006sec] 
\end{Soutput}
\end{Schunk}

The result is again an object of class~\class{evaluationResultList} for the
five recommender algorithms.
\begin{Schunk}
\begin{Sinput}
R> results
\end{Sinput}
\begin{Soutput}
List of evaluation results for 5 recommenders:

$`random items`
Evaluation results for 1 folds/samples using method ???RANDOM???.

$`popular items`
Evaluation results for 1 folds/samples using method ???POPULAR???.

$`user-based CF`
Evaluation results for 1 folds/samples using method ???UBCF???.

$`item-based CF`
Evaluation results for 1 folds/samples using method ???IBCF???.

$`SVD approximation`
Evaluation results for 1 folds/samples using method ???SVD???.
\end{Soutput}
\end{Schunk}

\begin{Schunk}
\begin{Sinput}
R> plot(results, ylim = c(0,100))
\end{Sinput}
\end{Schunk}
\begin{figure}
\centerline{\includegraphics[scale=1]{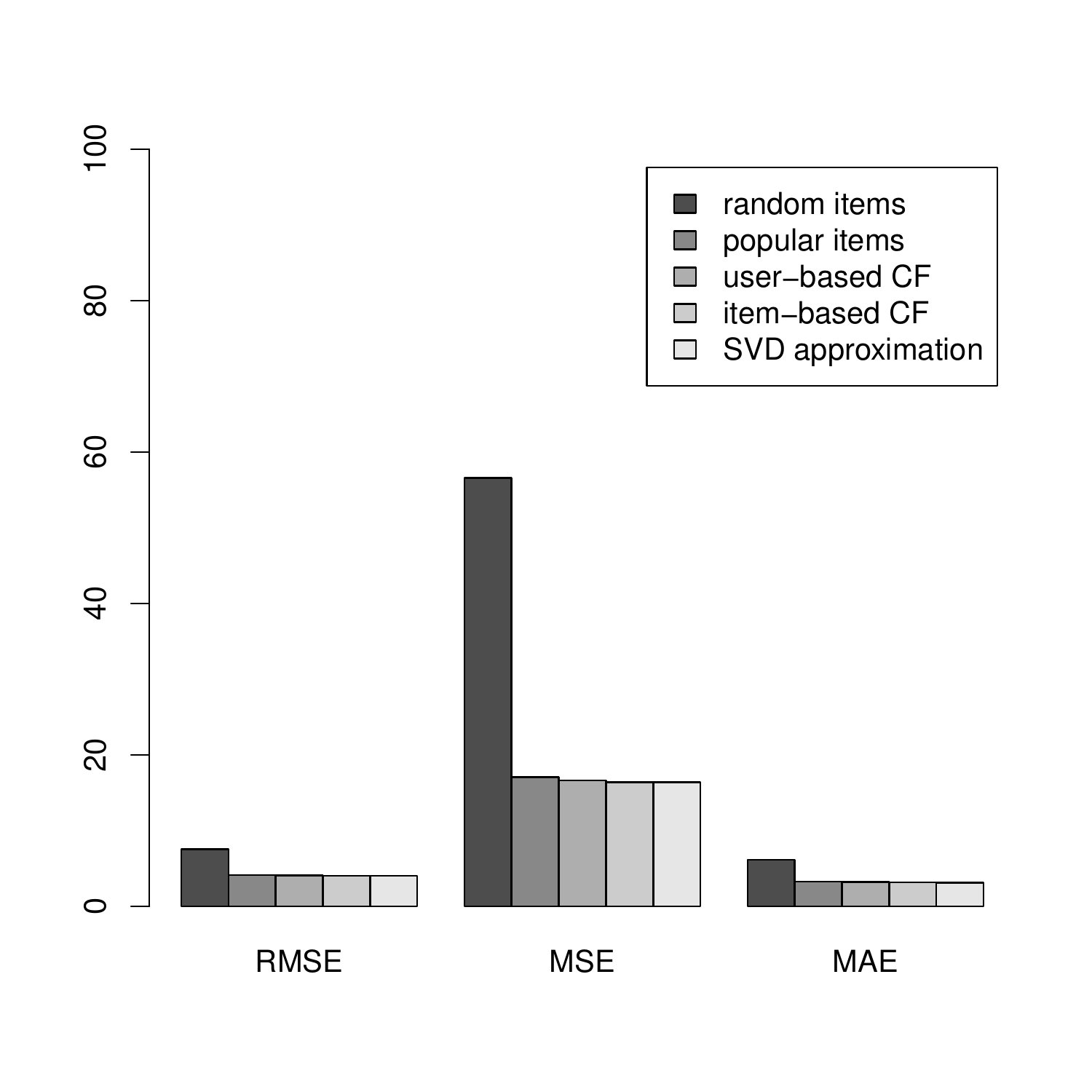}}
\caption{Comparison of RMSE, MSE, and MAE for recommender methods for the
given-3 evaluation scheme.}
\label{fig:real}
\end{figure}

Plotting the results shows a barplot with
the root mean square error, the mean square error and the mean absolute
error (see Figures~\ref{fig:real}).

\subsubsection{Using a 0-1 data set}
For comparison we will check how the algorithms compare given less information.
We convert the data set into 0-1 data and instead of a all-but-5 we use the given-3 scheme.

\begin{Schunk}
\begin{Sinput}
R> Jester_binary <- binarize(Jester5k, minRating=5)
R> Jester_binary <- Jester_binary[rowCounts(Jester_binary)>20]
R> Jester_binary
\end{Sinput}
\begin{Soutput}
1840 x 100 rating matrix of class ???binaryRatingMatrix??? with 67728 ratings.
\end{Soutput}
\begin{Sinput}
R> scheme_binary <- evaluationScheme(Jester_binary[1:1000],
+ 	method="split", train=.9, k=1, given=3)
R> scheme_binary
\end{Sinput}
\begin{Soutput}
Evaluation scheme with 3 items given
Method: ???split??? with 1 run(s).
Training set proportion: 0.900
Good ratings: NA
Data set: 1000 x 100 rating matrix of class ???binaryRatingMatrix??? with 36619 ratings.
\end{Soutput}
\end{Schunk}

\begin{Schunk}
\begin{Sinput}
R> results_binary <- evaluate(scheme_binary, algorithms,
+   type = "topNList", n=c(1,3,5,10,15,20))
\end{Sinput}
\begin{Soutput}
RANDOM run fold/sample [model time/prediction time]
	 1  [0.001sec/0.011sec] 
POPULAR run fold/sample [model time/prediction time]
	 1  [0.002sec/0.02sec] 
UBCF run fold/sample [model time/prediction time]
	 1  [0sec/0.29sec] 
IBCF run fold/sample [model time/prediction time]
	 1  [0.069sec/0.015sec] 
SVD run fold/sample [model time/prediction time]
	 1  
\end{Soutput}
\end{Schunk}

Note that SVD does not implement a method for binary data and is thus skipped.

\begin{Schunk}
\begin{Sinput}
R> plot(results_binary, annotate=c(1,3), legend="topright")
\end{Sinput}
\end{Schunk}
\begin{figure}
\centerline{\includegraphics[scale=1]{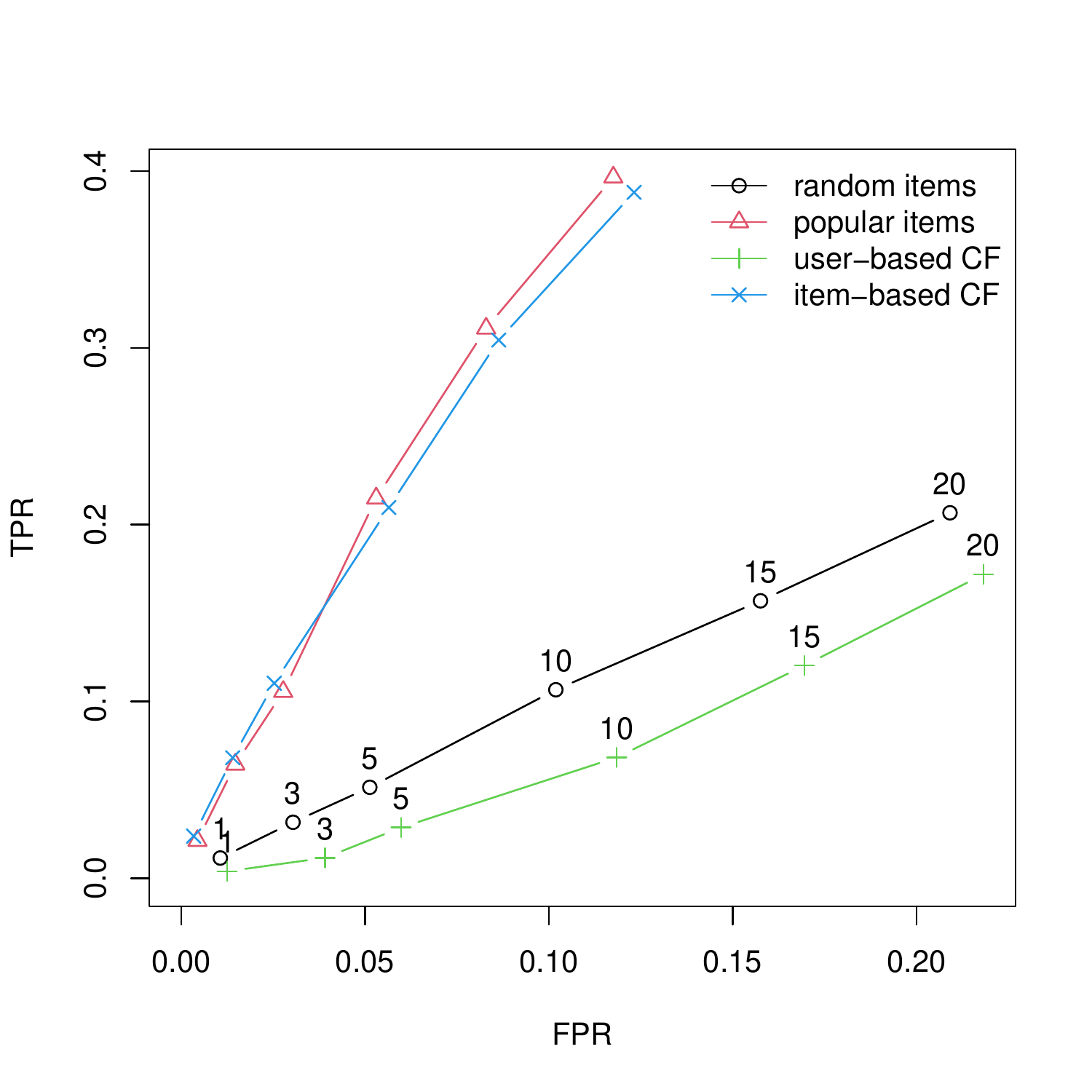}}
\caption{Comparison of ROC curves for several recommender methods for the
given-3 evaluation scheme.}
\label{fig:roc3}
\end{figure}

From Figure~\ref{fig:roc3} we see that given less information,
the performance of user-based CF suffers the most and the simple
popularity based recommender performs almost a well as item-based CF.

Similar to the examples presented here, it is easy to compare different
recommender algorithms for different data sets or to compare
different algorithm settings (e.g.,
the influence of neighborhood formation using different distance measures or different neighborhood sizes).


%
%
%
%
%
%
%
%
%


\subsection{Implementing a new recommender algorithm}

Adding a new recommender algorithm to \pkg{recommenderlab}
is straight forward since
it uses a registry mechanism to manage the algorithms.
To implement the actual recommender algorithm
we need to implement a creator function which takes a training data set,
trains a model and provides a predict function which uses the
model to create recommendations for new data.
The model and the predict function are both encapsulated in an
object of class~\class{Recommender}.

For examples look at the files starting with \code{RECOM} in the
packages \code{R} directory. A good examples is in \code{RECOM_POPULAR.R}.

\section{Conclusion}
\label{sec:conclusion}


%
In this paper we described the \proglang{R}
extension package~\pkg{recommenderlab}
which is especially geared towards developing and testing
recommender algorithms.
The package allows to create evaluation schemes following accepted
methods and then use them to evaluate and compare recommender algorithms.
\pkg{recommenderlab} currently includes several standard algorithms and
adding new recommender algorithms to the package is facilitated by the
built in registry mechanism to manage algorithms.
%
In the future we will add more and more of these algorithms to the package and
we hope that some algorithms will also
be contributed by other researchers.





\section*{Acknowledgments}
This research was funded in part by the NSF Industry/University Cooperative Research Center for Net-Centric Software \& Systems.

%
%

\bibliography{recommender,recom_talk,data,stuff,association_rules,recommenderlab_usage}

\end{document}